\documentclass[preprint,12pt]{elsarticle}

\usepackage{amssymb}
\usepackage{float} 
\usepackage{amstext} 
\usepackage{amsmath} 
\usepackage{graphicx}
\usepackage{subfigure} 
\usepackage{multirow}

\journal{arXiv}
\everymath{\displaystyle}
\begin{document}

\begin{frontmatter}


\title{Complete cavity map of the \emph{C. elegans} connectome}
\author[inst]{Bo Liu\fnref{fn1}}
\author[inst]{Rongmei Yang\fnref{fn1}}
\fntext[fn1]{These authors contributed equally to this work.}  
\author[inst2]{Hao Wang\fnref{fn1}}
\author[inst]{Linyuan L\"u\corref{cor}}
\cortext[cor]{linyuan.lv@uestc.edu.cn}
\affiliation[inst]{organization={Institute of Fundamental and Frontier Sciences, University of Electronic Science and Technology of China},
            city={Chengdu},
            postcode={610054}, 
            country={P. R. China}}

\affiliation[inst2]{organization={International Center for Complex Sciences, School of Science, Hainan University}, 
            city={Haikou},
            postcode={570228}, 
            country={P. R. China}}


\begin{abstract}
Network structure or topology is the basis for understanding complex systems. Recently, higher-order structures have been considered as a new research direction that can provide new perspectives and phenomena. However, most existing studies have focused on simple 1-cycles and triangles, and few have explored the higher-order and non-trivial cycles. The current study focused on the cavity, which is a non-trivial cycle. We proposed a method to compute cavities with different orders based on pruning in the boundary matrix and calculated all cavities of the neural network of Caenorhabditis elegans (\emph{C. elegans}) neural network. This study reports for the first time a complete cavity map of \emph{C. elegans} neural network, developing a new method for mining higher-order structures that can be applied by researchers in neuroscience, network science and other interdisciplinary fields to explore higher-order structural markers of complex systems.

\end{abstract}


\end{frontmatter}


\section{Introduction}

Complex networks, composed of nodes and edges, have become essential expressions for understanding complex systems in the real world\cite{barabasi2013network}. Recent years have seen a trend towards studying the higher-order (beyond pairwise) interactions in complex network.\cite{battiston2020networks,battiston2022higher}. Higher-order interaction refers to a simplex with multiple nodes (more than two nodes)\cite{bianconi2021higher} or a hyperlink in a hypergraph\cite{bretto2013hypergraph}. These interactions are crucial for understanding the structure and function of networks\cite{torres2021and}. Increasing evidence suggests that higher-order structures in systems can help us better understand and predict the dynamic behavior of complex systems\cite{iacopini2019simplicial,torres2020simplicial,battiston2021physics,millan2020explosive,shi2013searching}. 

The mathematical framework of algebraic topology can be used to study the structure of networks at a higher-order level, which can capture networks from different dimensions\cite{shi2019totally,shi2022simplicial,giusti2016two}. A simplex is a basic and vital structure in algebraic topology, also known as a clique. According to Poincaré's partitioning idea, a network can be considered as a complex composed of simplices\cite{zomorodian2004computing,hatcher2002algebraic}. Additionally, the topological space of the network can be further divided into a vector space composed of simplices with the same order as the basis. 

A higher-order (order $>$ 1) simplex is a cycle, such as a 2-simplex (triangle) is the smallest cycle\cite{shi2019totally}. The cycle is an important structure in complex networks\cite{fan2021characterizing}, however, previous studies have focused on simple structures such as triangles or 1-cycles. Combined with algebraic topology, we can explore higher-order cycles. An essential basis for cycle structures is to find nontrivial cycles, not only simplices. Bassett and colleagues\cite{sizemore2018cliques,sizemore2019importance} found cliques and cavities are abundant in the human brain connectome compared to null models, reflect the architecture of the brain for fast local processing, and play a significant role in some cognitive information processes and behaviors.

In addition to exploring cliques in a network, finding cavities (nontrivial cycles) that are surrounded by cliques with the same order is more challenging \cite{reimann2017cliques}. A cavity is defined as the shortest non-clique cycle of independent equivalence classes\cite{shi2022simplicial}. A previous study \cite{shi2021computing} proposed a method to compute cavities in neural networks based on 0-1 programming to find all 3-cavities and a few 1-cavities and 2-cavities for \emph{Caenorhabditis elegans} (\emph{C. elegans})\cite{C.elegans}. However, there is no efficient, accurate, and fast method for calculating all higher-order structures.

Here, we propose a method to compute all cavities for a given network, which can calculate cavities at any order. The method consists of three steps. First, we search for a spanning tree and all cliques. Second, we find linearly independent cycles based on the pruning method in the boundary matrix. Third, we analyse the hollow cycles and calculate all cavities. To our knowledge, this is the first report of a complete cavity map in the neural network of \emph{C. elegans}.

\section{Results}

A \emph{k}-cavity is a $k$-cycle in a linearly independent equivalence class of the homology group $H_k$, which must be hollow and surrounded by \emph{k}-cliques. The essence of searching for cavities is the process of searching for independent equivalence classes for kernel space $Z_k$ through image space $Y_k$. We first need to identify the independent cycles in $Z_k$, and the problem is transformed into finding a \emph{k}-order spanning tree in $m_k$ \emph{k}-cliques, composed of $r_k$ \emph{k}-cliques. Then we add the remaining ($m_k-r_k$) \emph{k}-cliques in turn to the spanning tree and then use the pruning method to obtain a set of independent \emph{k}-cycles. For these independent cycles, we aim to find hollow cycles, which we consider to be the shortest of the cycles that are equivalent to them. Finally, for all the hollow cycles, we verify that the number of independent equivalence classes is equal to the Betti number. A detailed description of this process is provided below.

\subsection{Searching a spanning tree}

To find the cavities in a network, we start from find the independent cycles via a spanning tree. The maximal linearly independent group of the boundary matrix $B_k$ corresponding to the $k$-order spanning tree corresponding to the network $G$. There are many ways to find the maximally linearly independent group in a matrix. Here, we simplify the boundary matrix into row-echelon form and select the column vector corresponding to the position of the echelon as the maximal linearly independent group.

\begin{figure}[H]
\centerline{\includegraphics[width=14 cm]{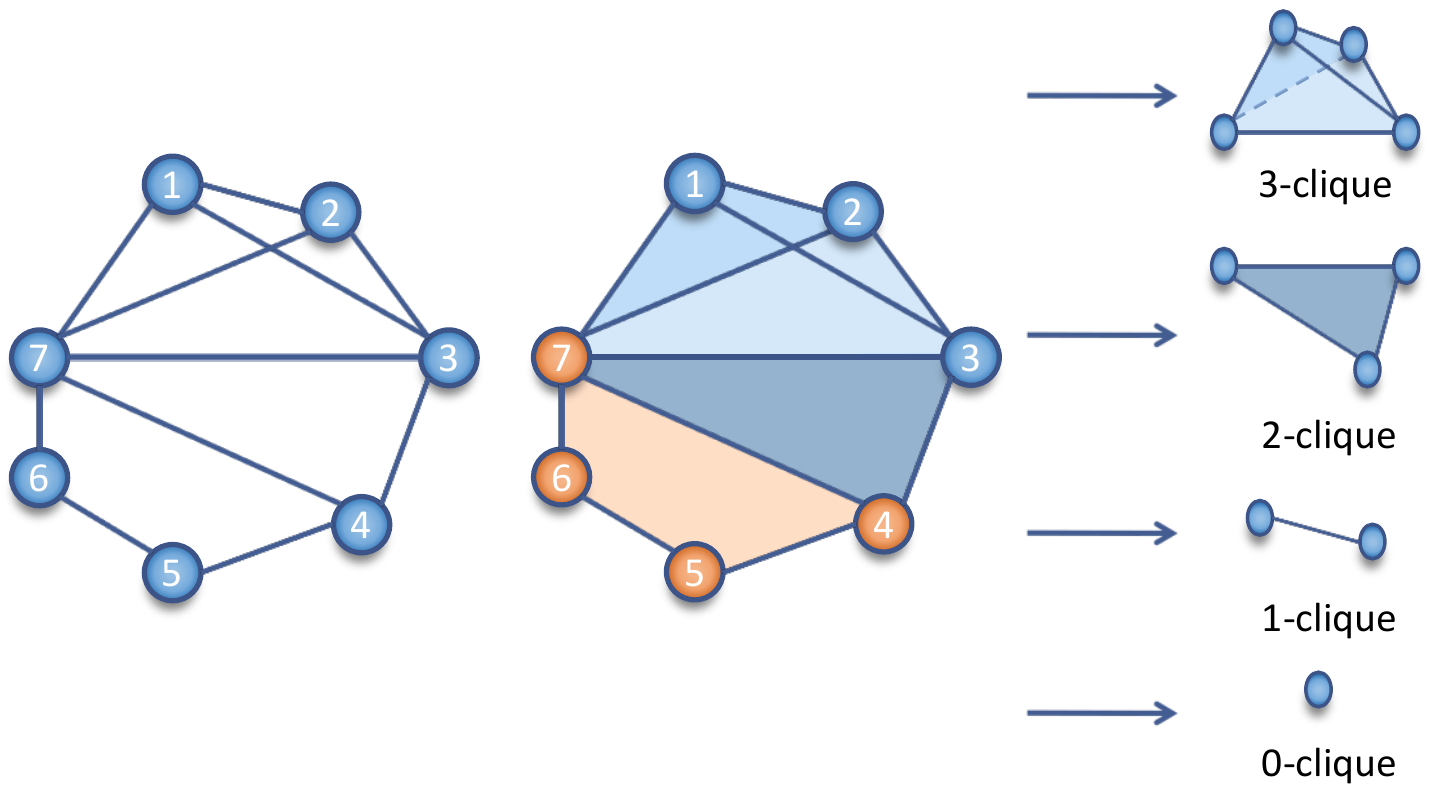}}
\caption{(Color online) \textbf{A simple network with seven nodes and eleven edges.} a. a pairwise network composed of nodes and edges. b. a simplicial network include seven 0-cliques, eleven 1-cliques, five 2-cliques and one 3-clique.
\label{fig1}}
\end{figure}

\begin{figure}[H]
\centerline{\includegraphics[width=14 cm]{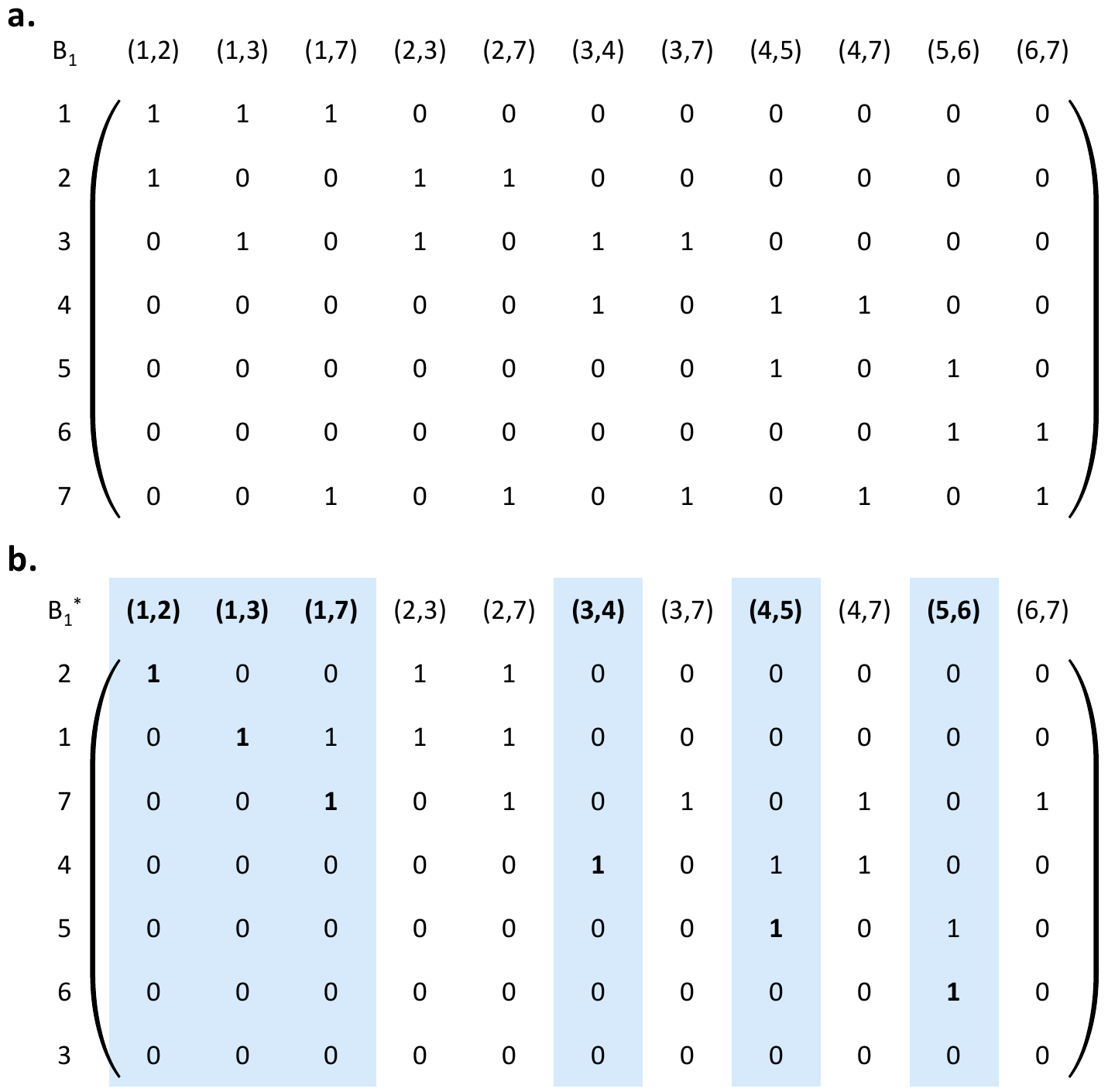}}
\caption{(Color online) \textbf{The computation process of a spanning tree in the toy network.} a. The one order boundary matrix $B_1$ of the toy network, columns denotes all 1-cliques, row denotes all 0-cliques. b. $B_1^*$ is reduced row echelon form of matrix $B_1$, the first non-zero element in each row is at the echelon, and corresponding columns are a maximal linearly independent system of that matrix. The 1-cliques corresponding to these columns (shaded columns) form the one order spanning tree of that network. 
\label{fig2}}
\end{figure}

Take computing a one-order spanning tree for the toy network (see figure~\ref{fig1}) as an example, the calculation process is shown in Figure~\ref{fig2}. We obtain the simplified matrix $B_1^*$ from the boundary matrix $B_1$. And then select the column vectors at the echelon (first nonzero position in each row, corresponding to the column vectors in bold in the $B_1^*$). Lastly, the 1-cliques that correspond to the one-order spanning tree are (1,2), (1,3), (1,7), (3,4), (4,5), and (5,6). The spanning tree in a network is not unique, and in this paper we find a spanning tree with the largest weight based on the weighted \emph{C.elegance} neuron network. We define the weight of each column is defined as the sum of the edge weights of the clique corresponding to that column. It is only necessary to arrange the columns in descending order of weights before simplifying the boundary matrix. In this way, we can find a unique spanning tree in a network and ensure that the subsequent independent cycles generated based on this are made up of higher-weight cliques.

\subsection{Linearly independent cycles based on pruning in boundary matrix}

If we have found a $k$-order spanning tree, the remaining $k$-cliques can be placed into the spanning tree to create a set of linearly independent cycles. Previous studies have used optimization methods for determining constraint conditions to determine cycles, which can be complex, and here we propose a method for pruning in the boundary matrix. The most crucial process in pruning is to identify leaf cliques that are not involved in the formation of a cycle. It can be thought that a 1-cycle is composed of 1-cliques (edges), and each node of a 1-cycle is connected to two 1-cliques (edges), a 2-cycle is composed of 2-cliques(triangles), and each edge of a 2-cycle is connected to two 2-cliques(triangles). By that analogy, a $k$-cycle is composed of $k$-cliques, and each ($k-1$)-clique is connected to two \emph{k}-cliques. A leaf clique, which corresponds to rows whose row values add up to 1 in a boundary matrix, is the \emph{k-1}-clique is connected 1 $k$-cliques. After identifying the leaf cliques in the matrix, we deleted the corresponding rows and columns in the matrix. This process is iterative, after each deletion of columns means that new leaf cliques may be created in the  structure to be simplified. The deletion continues until there are no leaf cliques in the structure to be simplified, resulting in a cycle.

In the sample network (see figure 1), we add the clique (6,7) to the one-order spanning tree. First, we sum the boundary matrix of the structure by each row, and we find leaf node 2 whose row sum is less than two. Then, the column and row of node 2 are deleted (see step1 matrix), and the new matrix is obtained (see step2 matrix). Finally, we check no leaf nodes in a new matrix and find a cycle ([(1,3), (1,7), (3,4), (4,5), (5,6), (6,7)]) corresponding to the new added clique (6,7).

\subsection{Hollow cycles in linearly independent cycles}

The structures obtained by pruning are a set of independent \emph{k}-cycles. However, since empirical networks are complex, there may be the following three cases to further determine this cycle: (1) a ($k+1$)-clique; (2) a \emph{k}-cavity; (3) a \emph{k}-cycle that is not the shortest cycle in the equivalence class or the sum of multiple \emph{k}-cavities. To avoid the first case, limit the length of the \emph{k}-cavity to greater than the shortest length $2^{k+1}$. If we distinguish between the second or the third case of a \emph{k}-cycle, the cycle should be evaluated for one important aspect: whether it is hollow. We consider that only a hollow cycle will not be reduced, that is, it does not contain a cycle shorter than it.

The next step is to determine whether a cycle of the set of linearly independent cycles is a hollow cycle. This \emph{k}-cycle can be decomposed into several 0-cliques. Next, we filter through all the $k$-cliques to find the ones that contain all the 0-cliques above. If the number of filtered $k$-cliques is the same as the number of $k$-cliques that make up the \emph{k}-cycle, then the \emph{k}-cycle is a hollow cycle if no further judgement is needed. We extract the 0-cliques, from the cycle of ([(1,3), (1,7), (3,4), (4,5), (5,6), (6,7)]) found in section 2.2, which are [1,3,4,5,6,7]. Furthermore, we found the 1-cliques formed by those 0-cliques ([(1,3), (1,7), (3,4), (4,5), (5,6), (3,7), (4,7), (6,7)]), since the filtered 1-cliques have two new ones, the cycle is not a hollow cycle and needs further determination. 

For the filtered structure consisting of filtered \emph{k}-cliques, We construct all the 1-cliques in this structure as a network $G^{'}$, if it's ${\beta_{k}^{G^{’}}=1}$, we first simplify the ($k+1$)-cliques in it. The $k$-cliques of each  ($k+1$)-cliques are compared with the \emph{k}-cliques in the original \emph{k}-cycle. We calculate the number of repeated \emph{k}-cliques in the two structures, denoted as $N_r$. Since there are $k+2$ \emph{k}-cliques in each ($k+1$)-clique, the number of remaining non-repeating \emph{k}-cliques is $N_nr = k-N_r+2$. And then we need to reserve the smaller of the number of repeated cliques and non-repeated cliques in this ($k+1$)-clique. There may be more than one ($k+1$)-cliques in the filtered structure, and we give priority to the one with the larger $r$. Simplification is a dynamic iterative process that updates the $(k+1)$-clique in the structure after a $(k+1)$-clique is processed. After simplifying the structure we need to calculate the row sums of the boundary matrix corresponding to this structure, if each row sum is equal to 2, then the structure is a hollow cycle. For a filtered structure, we only consider two possible cases, one where ${\beta_{k}^{G^{’}}>1}$ is out of our discussion, and the other ${\beta_{k}^{G^{’}}=1}$ where the other ($k+1$)-clique need to be simplified first, such as the 2-clique (1,2,5) in the second figure of figure~\ref{fig4}a. In this structure $N_r = 2 >1$ and $N_nr = 1$, we need to reserve the  non-repeated cliques in this ($k+1$)-clique. We obtain a new structure [(2,3),(2,5),(3,4),(4,5)], which is a hollow cycle and contain no higher-order cliques.   
Figure~\ref{fig4}b shows the simplification process for a non-hollow 2-cycle. When the structure are filtered, new clique (1,2,3) are created. Both filtered structures have a higher-order clique, compared to the original 2-cycle, $N_r=2$ and 3, $N_nr=1$. Therefore, the non-overlapping clique (1,2,3) is retained. The case $N_r=N_nr=2$ may also occur in the 2-cycle, as shown in figure~\ref{fig4}b, we only remain one of the two types cliques (the repeating and non-repeating cliques) and delete the other.

Returning to the non-hollow structure we restored above in the example network is shown in figure~\ref{fig5}c. It can be seen that the restored structure has two new 1-cliques,(3,7)and (4,7). We form a network $G^{'}$, which with all recovered all 1-cliques, ${\beta_1^{G^{'}}}=1$. Next, we extract the 2-cliques, which are (1,3,7) and (3,4,7), $N_r(1,3,7)=2$ and $Nr_(3,4,7)=1$. Since $N_nr(1,3,7)=2 > N_r(3,4,7)=1$, we first simplify (1,3,7). And then $N_r(1,3,7)>N_nr(1,3,7)$, We retain the 1-clique (3,7) and delete the repeat 1-cliques (1,7) and (1,3). We update the 2-cliques in this structure again, only (3,4,7). At this time $N_r(3,4,7)=2$,$N_nr(3,4,7)=1$. We retain the edge (4,7) and delete the repeat edges (1,7) and (1,3). There is no higher-order clique (2-clique) in this structure, and all row sums of its boundary matrix are equal to 2 as shown in figure~\ref{fig6}a. Therefore, it is a hollow 1-cycle.   

For the structure consisting of filtered \emph{k}-clqiues, if it is ${\beta_{k}^{G^{’}}}>1$. The next step is to find a spanning tree in $G^{'}$, find the linearly independent cycles in the spanning tree, further find the hollow cycles in the filtered linearly independent cycles, We only simplify structures with Betti numbers equal to one, and save the simplified hollow cycles. If the number of independent equivalence classes of all hollow cycles equals $\beta_{k}^{G^{’}}$, the calculation stops. If it is less than $\beta_{k}^{G^{’}}$, the computation continues by finding a spanning tree in $G^{'}$ again. 

\subsection{Cavities in hollow cycles}

Three steps were required to find the cavities in all the hollow circles. The first step is to transform all hollow circles into column vectors, the i-th hollow cycle can be changed into $X^{T}_{i}=(m_1,m_2, ... ,m_k)^T$, $m_i=0$ or 1, if the i-th $k$-clique is a face of the $k$-cavity $m_i=1$, otherwise 0. The second step is to arrange these column vectors in ascending order according to the length of the hollow cycles, if the cycles with the same length and larger weights correspond to the column vectors in front. The final step reduces this matrix to a row ladder matrix taking out the number of hollow cycles corresponding to each ladder, the number of which is $\beta_k$, and these hollow cycles are all the $k$-cavities in this network.

As can be seen from the matrix in figure~\ref{fig6}b, the hollow cycle found in Figure 5c is represented by the matrix $x^{T}$, and since there is only one cycle, there is only one column in $x^{T}$. Combining the matrix with the higher-order boundary matrix ($B_2$) gives $rank(x^{T},B_2) - rank(B_2) = 1$, which is equal to the 1st order Betti number $\beta_{1}^{G}$ of the network, so this hollow cycle is the 1-cavity of this network.

\section{Conclusion}

To conclude, we present a novel method to calculate the cavities with different orders. We solve the problem that the spanning trees are not unique by using weights to identify a unique spanning tree in a network. Subsequently, a pruning algorithm is proposed to find the cavities. Our study is the first to report a whole cavity map of \emph{C. elegans} neural network. Recent studies have demonstrated that nontrivial higher-order structures play an essential role in brain connectomes, neural circuits, and other complex systems. Due to the size of these complex systems, fast computing for cavities of larger networks is urgently needed. Even edges usually contain direction information in biological systems, so it is worth considering defining cliques, cavities, and boundary matrices in directed networks.

\section{Materials and method}
\subsection{Data description}
\emph{C. elegans} neural network is a model organism in developmental biology and neurological research. In this study, we use a weighted \emph{C. elegans} neural network with 297 nodes and 2148 edges\cite{achacoso2022ay}. Nodes represent neurons, an edge indicates a synapse between two neurons, and weight denotes the number of synapses. We transform the original directed data into undirected. If there is only one edge between two nodes, we keep the weights and remove the direction. If there are multiple edges between two nodes, the weights are summed, and the direction is removed. The number of $k$-cliques $m_k$ and the number of $k$-cavities $\beta_k$ are shown in Table \ref{table1}. The visualization of complete cavity map is shown in Supplementary material. We use different shapes and colors to represent the cell type corresponding to the node, where the orange sphere represents the interneuron, the blue cube denotes the motorneuron, the green octahedron is the sensory, and the brown is other type.

\subsection{Clique}
A clique, which is a fully connected structure, is called simplex in algebraic topology.  A clique of order \emph{k}, called the \emph{k}-clique. In this study, we define a \emph{k}-clique, a fully connected structure with $k+1$ nodes. For example, a node is a 0-clique, an edge is a 1-group, a triangle is a 3-clique, a tetrahedron is a 4-clique, etc. It is well known that a structure consisting of simplices is called a complex. Binary networks can be decomposed into cliques (simplex) as a unit, so such networks can be called a complex, the maximum order of the simplex of a network is the order of this network. 

\subsection{Vector space}

Cliques can be used as the unit of a network. The most common topological structure composed of nodes and edges can be regarded as the structure under the perspective of 1-cliques. In linear algebra, it can also be regarded as the structure in the vector space $C_1$ consisting of 1-cliques. Similarly, it can also be extended to construct the vector space $C_k$, in which all \emph{k}-cliques in this vector space are basis vectors, and the dimension is $m_k$ the number of \emph{k}-cliques, \emph{k}-chains composed of \emph{k}-cliques are elements in $C_k$. 

\subsection{Boundary operator and boundary matrix}

A clique can be understood as a cycle formed by its lower-order cliques, which means that given a \emph{k}-clique ($k>0$), it can be decomposed into lower order cliques. Defining a boundary operator is an operation method from the higher-order vector space to the lower-order vector space. The boundary operator $\partial_k$ can operate on a $k$-order vector Space $C_k$, and the relation is $C_{k} \stackrel{\partial_{k}}{\rightarrow} C_{k-1}$. The boundary matrix $B_k$ can be further defined by the relation between the \emph{k}-cliques and its boundary, which is the matrix expression of the boundary operator. The columns in the boundary matrix $B_k$ represent the $k$-cliques and the rows represent the ($k-1$)-cliques, where each element represents whether the corresponding ($k-1$)-clique is a face of the $k$-clique (the ($k-1$)-clique is a part of the \emph{k}-clique). 

\subsection{Homology}
The elements of the vector space $C_k$ are $k$-chains, and these chains form a chain group. Where the closed chain $l$ forms the closed chain group $Z_k$, which is $\partial_{k}(l)=0$, this is reminiscent of the concept of kernel space, $Z_k=\operatorname{ker} \partial_{k}$. Higher-order vector spaces can also be mapped by boundary operators, $\ldots \rightarrow C_{k+1} \stackrel{\partial_{k+1}}\rightarrow C_{k} \stackrel{\partial_k}\rightarrow C_{k-1} \rightarrow \ldots $ and thus boundary groups are generated, $Y_{k}=im\partial_{k+1}$. Because there is another important property of the boundary operator $\partial_{k}\partial_{k+1}=0$, so $Y_{k} \subseteq Z_{\mathbf{k}}$, and then $Y_{k} \subseteq Z_{k} \subseteq C_{k}$.

If there are two $k$-cycles $l_i$ and $l_j$ in a network, if $l_i+l_j$ is the boundary of a ($k+1$)-chain, $l_i$ and $l_j$ are equivalent and belong to an equivalence class. All cycles in $Z_k$ can be divided into equivalence classes using $Y_k$, so the \emph{k}-th homology group in a network can be defined as $H_{k}=Z_{k} / Y_{k}$, the elements in $H_k$ are these equivalence classes (see figure~\ref{fig7}). The rank of $H_k$ is the number of cavities, which is Betti number, one of the important topological invariants. It can be deduced from the relations of vector spaces that the rank of $H_k$ is equal to the rank of $Z_k$ ($m_k-r_k$) minus the rank of $Y_k$ ($r_{k+1}$), so $\beta_k=m_k-r_k-r_{k+1}$. 


\unskip

\unskip

\begin{figure}[H]
\centerline{\includegraphics[width=14 cm]{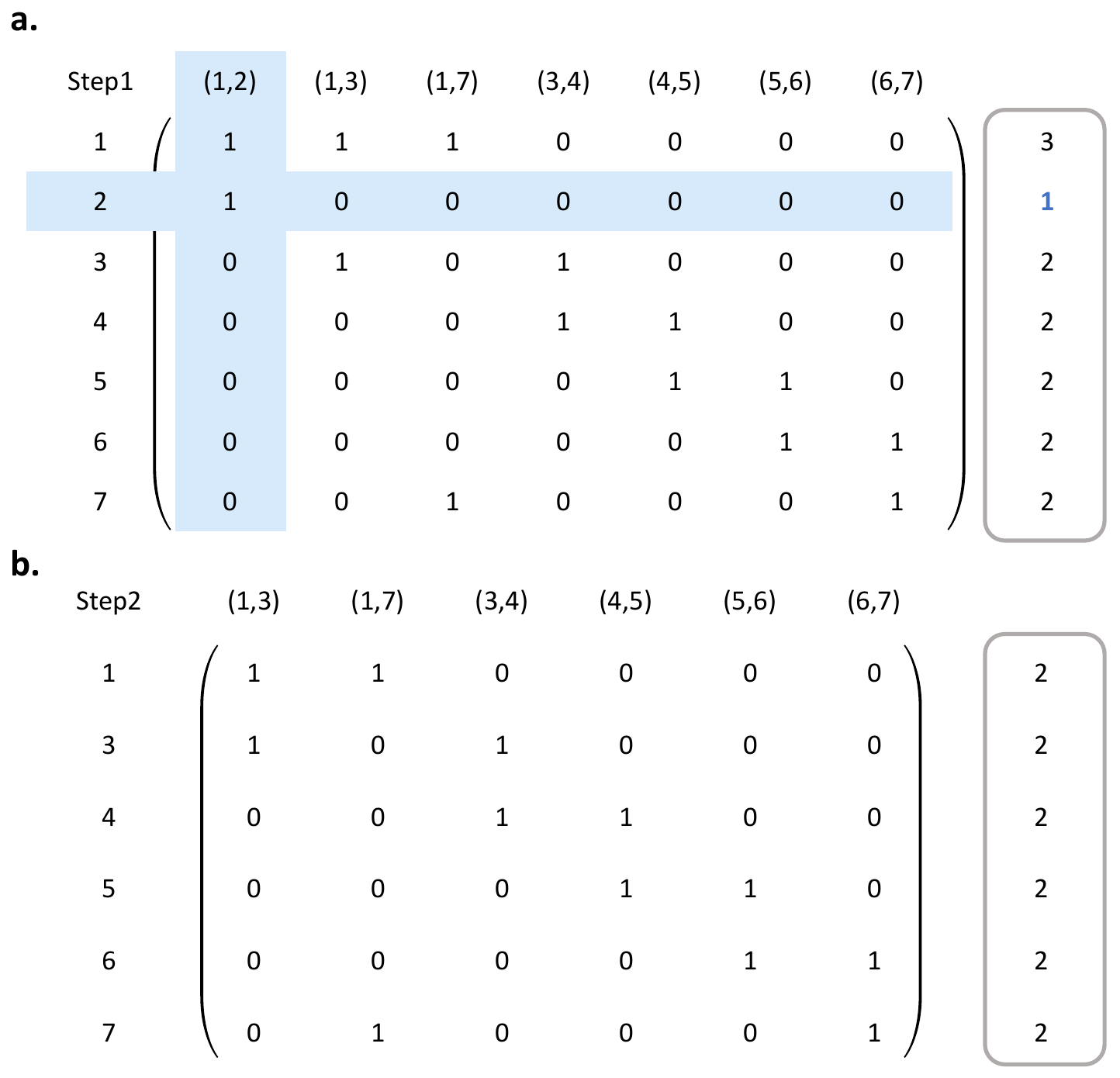}}
\caption{(Color online) \textbf{Pruning of the simple network.} a. The boundary matrix of the structure corresponding to the clique (6,7) add to the  1-spanning tree (see section 2.1). On the far right are the row sums, we have selected the leaf clique(node 2) and deleted their corresponding row and column. b. The boundary matrix corresponds to the structure after the deletion of the leaf clique in the first step. It was tested for leaf cliques and all the rows and sums in this boundary matrix are equal to two. There were no leaf cliques and it is determined that this structure is a 1-cycle.
\label{fig3}}
\end{figure}   
\unskip

\begin{figure}[H]
\centerline{\includegraphics[width=14 cm]{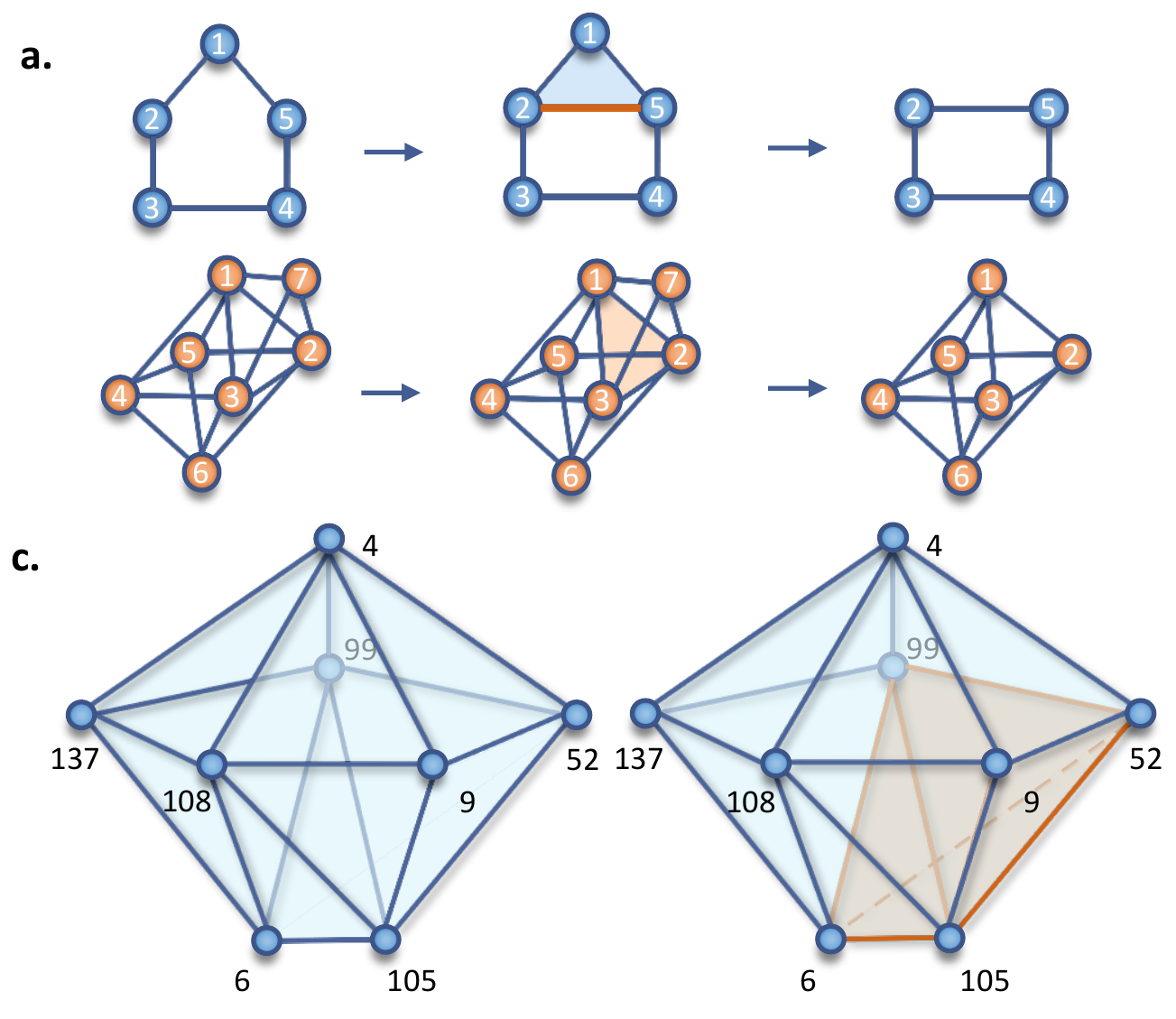}}
\caption{(Color online) \textbf{Simplifying the filtered cycle into an independent cycle by pruning.} a. The first row is the simplification of the hollow 1-cycle, the second row is the simplification of the hollow 2-cycle. In each row, first cycle is one of a set of linearly independent cycles, the second is the filtered structure of the first cycle, the third is the simplified hollow cycle of the second structure. b. The case when the higher-order clique in the filtered structure(right) is the same as the repeating and non-repeating cliques that composed it compared with the original cycle(left). 
\label{fig4}}
\end{figure}   
\unskip

\begin{figure}[H]
\centerline{\includegraphics[width=14 cm]{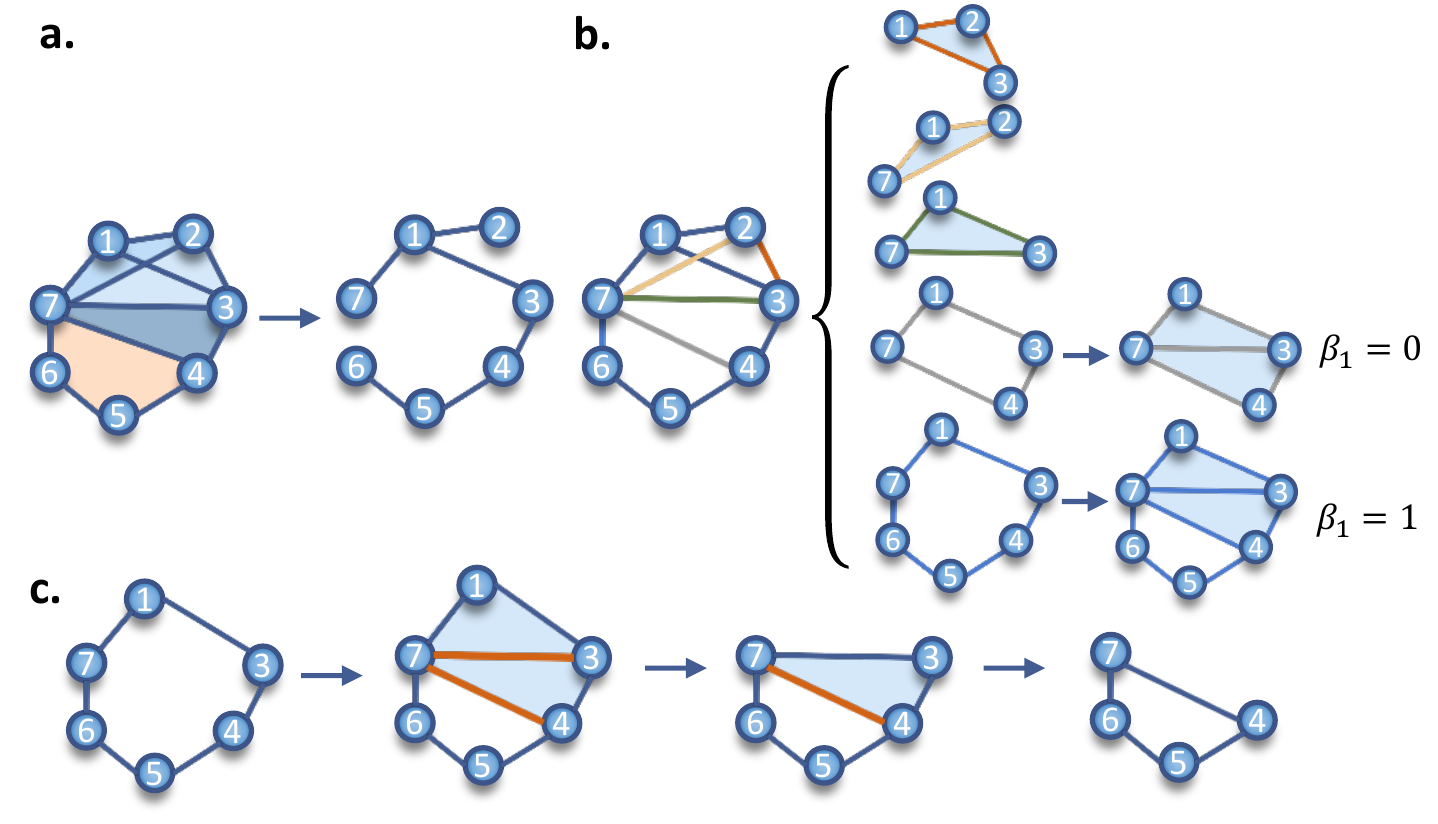}}
\caption{(Color online) \textbf{The process of finding hollow 1-cycles in the toy network(see Figure~\ref{fig1}).} a. The 1st spanning tree of the toy network. b. A set of linearly independent 1-cycles. c. The process of reducing unrelated cycles with Betti number equal to 1 to a hollow cycle.  
\label{fig5}}
\end{figure}   
\unskip

\begin{figure}[H]
\centerline{\includegraphics[width=14 cm]{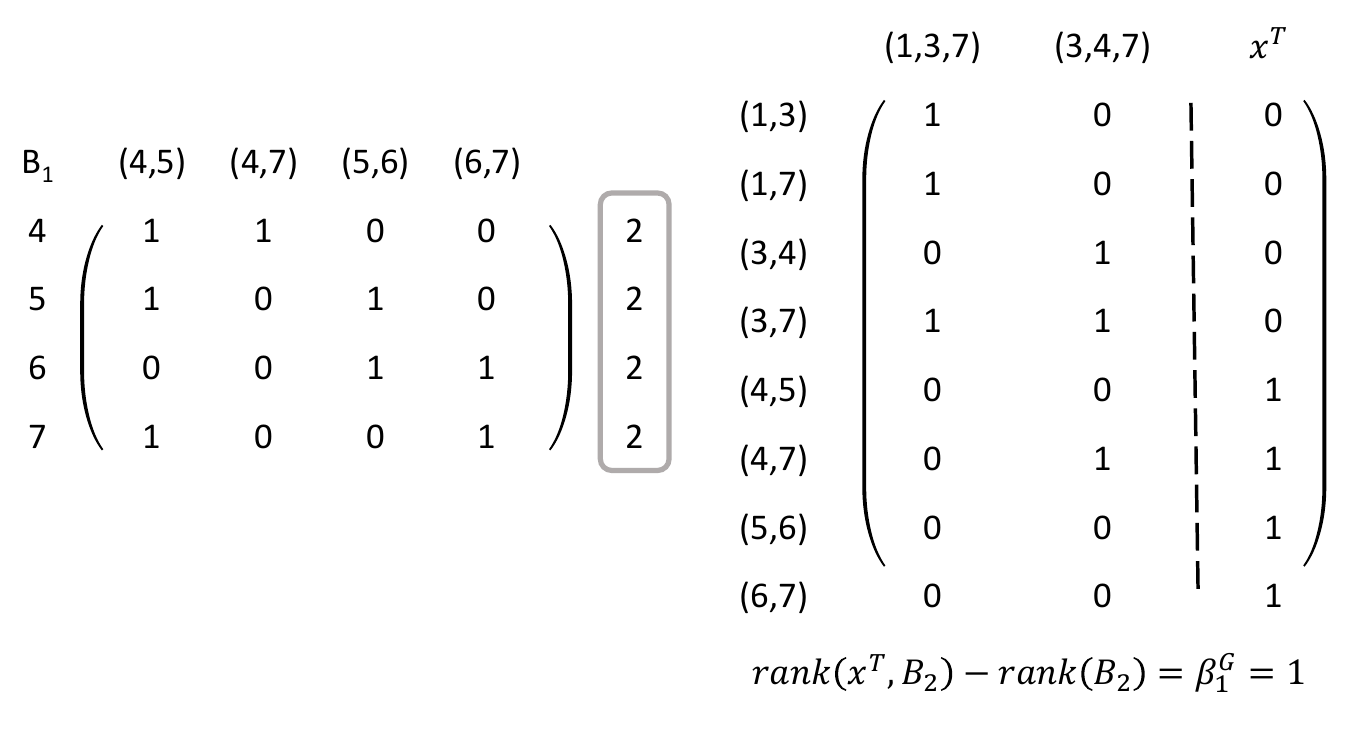}}
\caption{(Color online) \textbf{Determining if the hollow cycle of Figure~\ref{fig5} is a cavity by the boundary matrix.} a. The boundary matrix of the hollow 1-cycle (4,5,6,7) of Figure~\ref{fig5}c. b. A matrix consisting of a hollow cycle as a vector and a higher-order boundary matrix merged.
\label{fig6}}
\end{figure}   
\unskip

\begin{figure}[H]
\centerline{\includegraphics[width=14 cm]{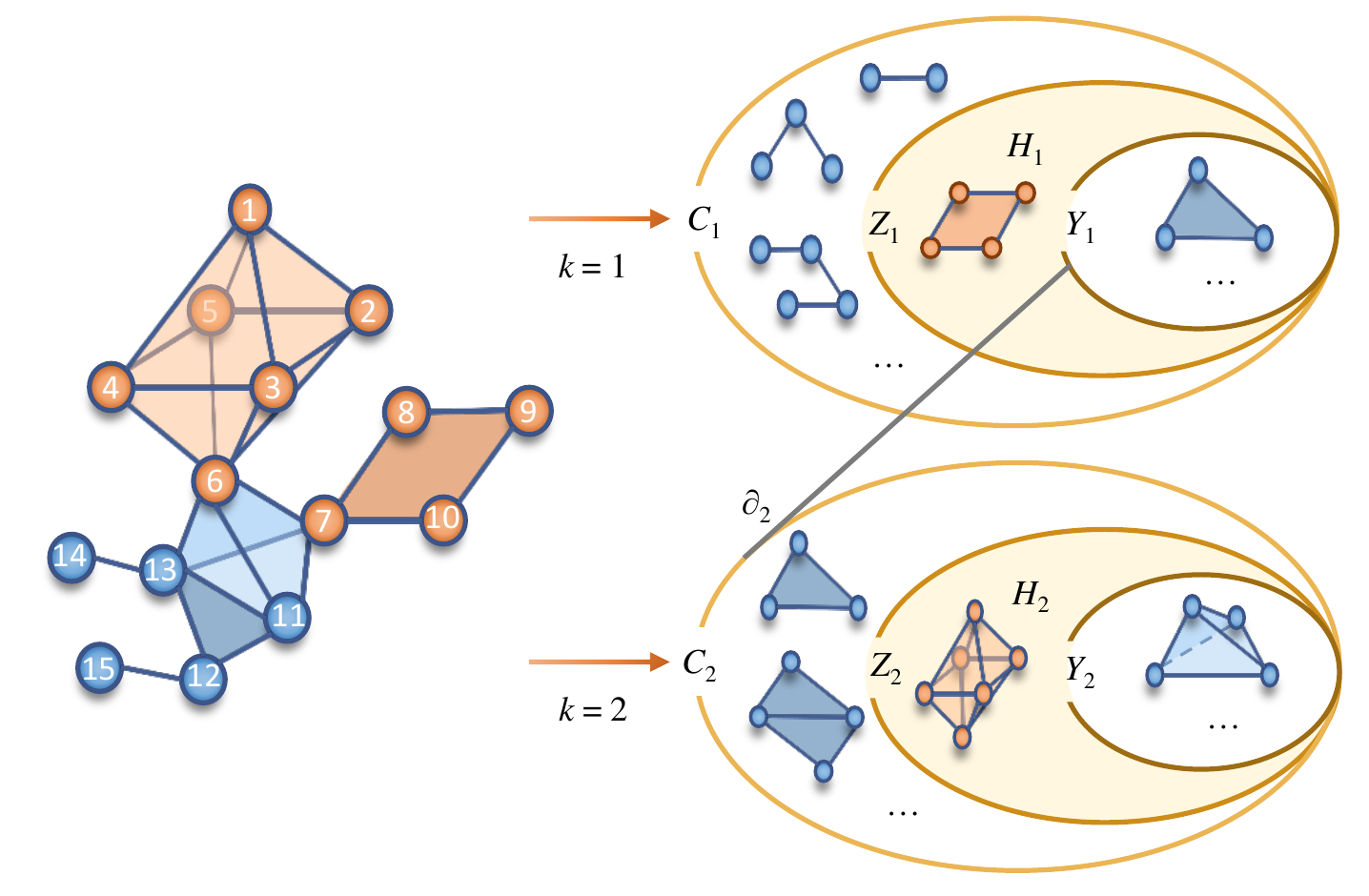}}
\caption{(Color online) \textbf{The relationship between a simple network and its adjacent vector spaces and the boundary operator.} We provide a network with ten nodes and its vector space with k = 1 and k = 2.
\label{fig7}}
\end{figure}   
\unskip

\begin{table}[H]
\centering
\caption{The number of cliques and cavities in \emph{C. elegans}  neural network}
\begin{tabular}{llllllllll} 
\hline
\multirow{2}{*}{Clique} & $m_0$    & $m_1$    & $m_2$    & $m_3$    & $m_4$   & $m_5$  & $m_6$ & $m_7$ & $m_8$  \\ 
\cline{2-10} & 297   & 2148  & 3241  & 2010  & 801   & 240 & 40 & 2  & 0   \\ 
\hline
\multirow{2}{*}{Cavity} & $\beta_0$ & $\beta_1$ & $\beta_2$ & $\beta_3$ & $\beta_4$ & $\beta_5$  & $\beta_6$ & $\beta_7$ & $\beta_8$  \\ 
\cline{2-10}  & 1     & 139   & 121   & 4     & 0     &   0 &  0 & 0  & 0  \\
\hline
\end{tabular}
\label{table1}
\end{table}
\unskip


 \bibliographystyle{elsarticle-num} 
 \bibliography{reference}

\section*{Supplementary material}
\subsection*{1-cavity map}
\begin{figure}[H]
\centerline{\includegraphics[width=0.95\linewidth]{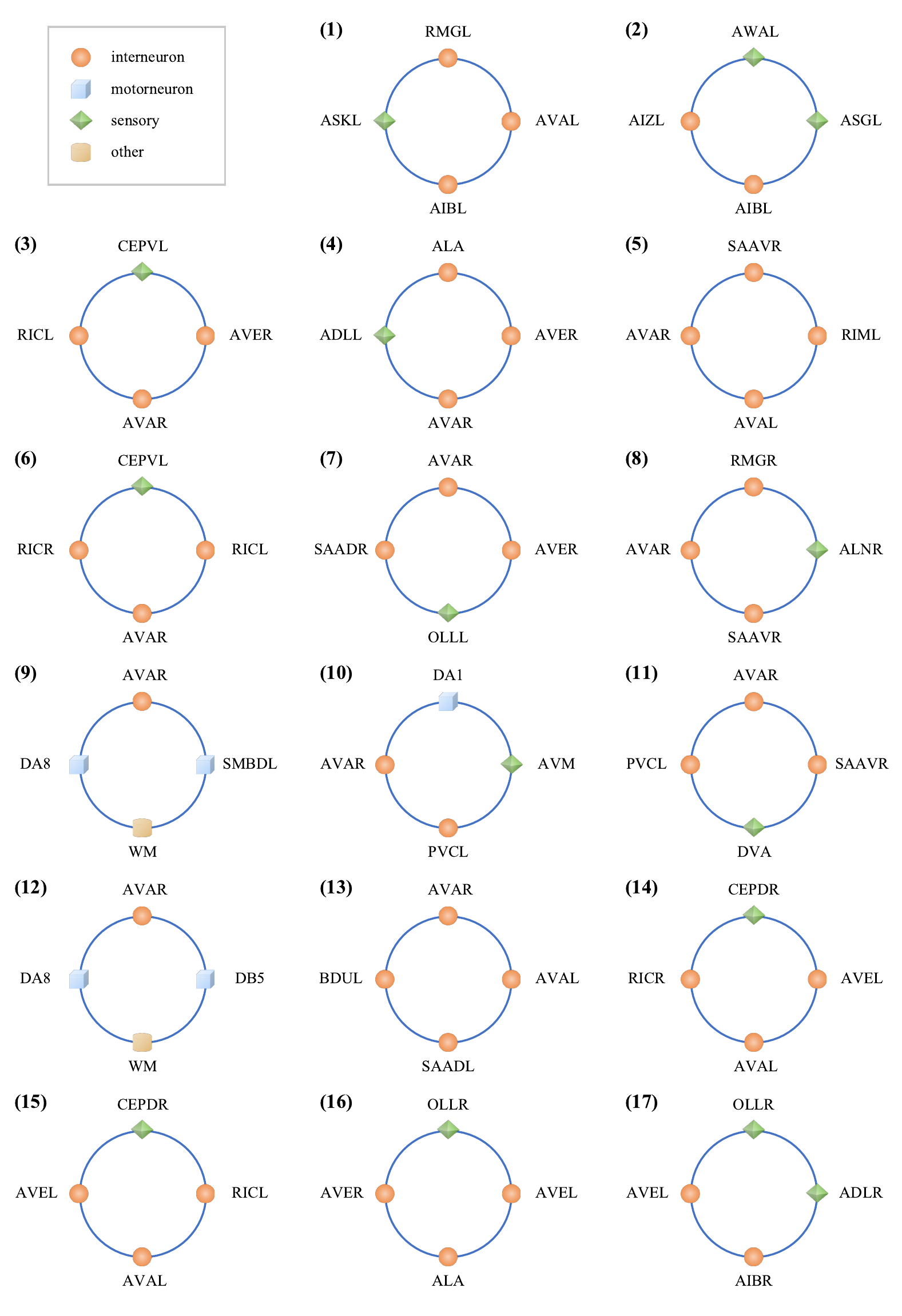}}
\end{figure}
\begin{figure}[H]
\centerline{\includegraphics[width=0.95\linewidth]{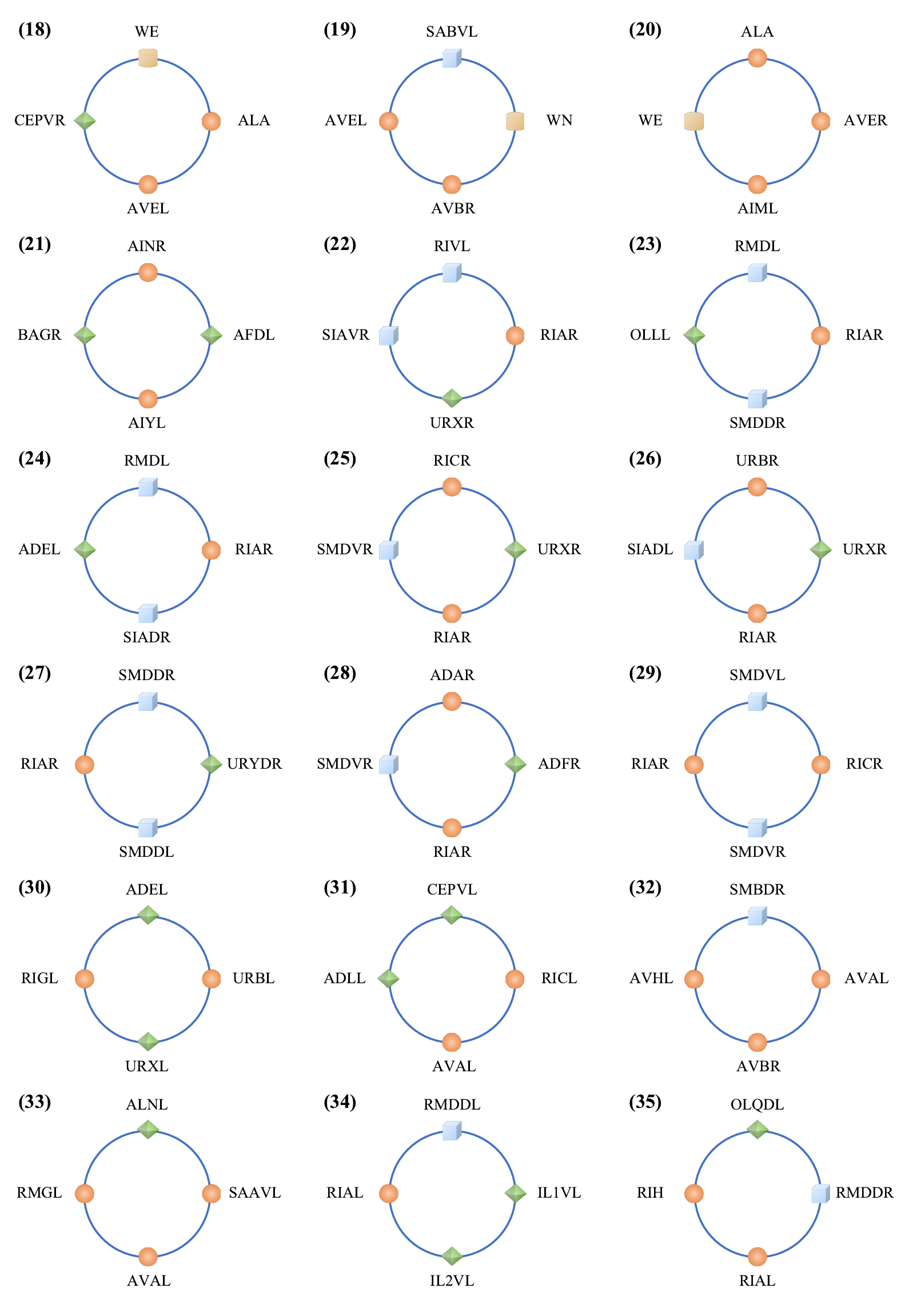}}
\end{figure}
\begin{figure}[H]
\centerline{\includegraphics[width=0.95\linewidth]{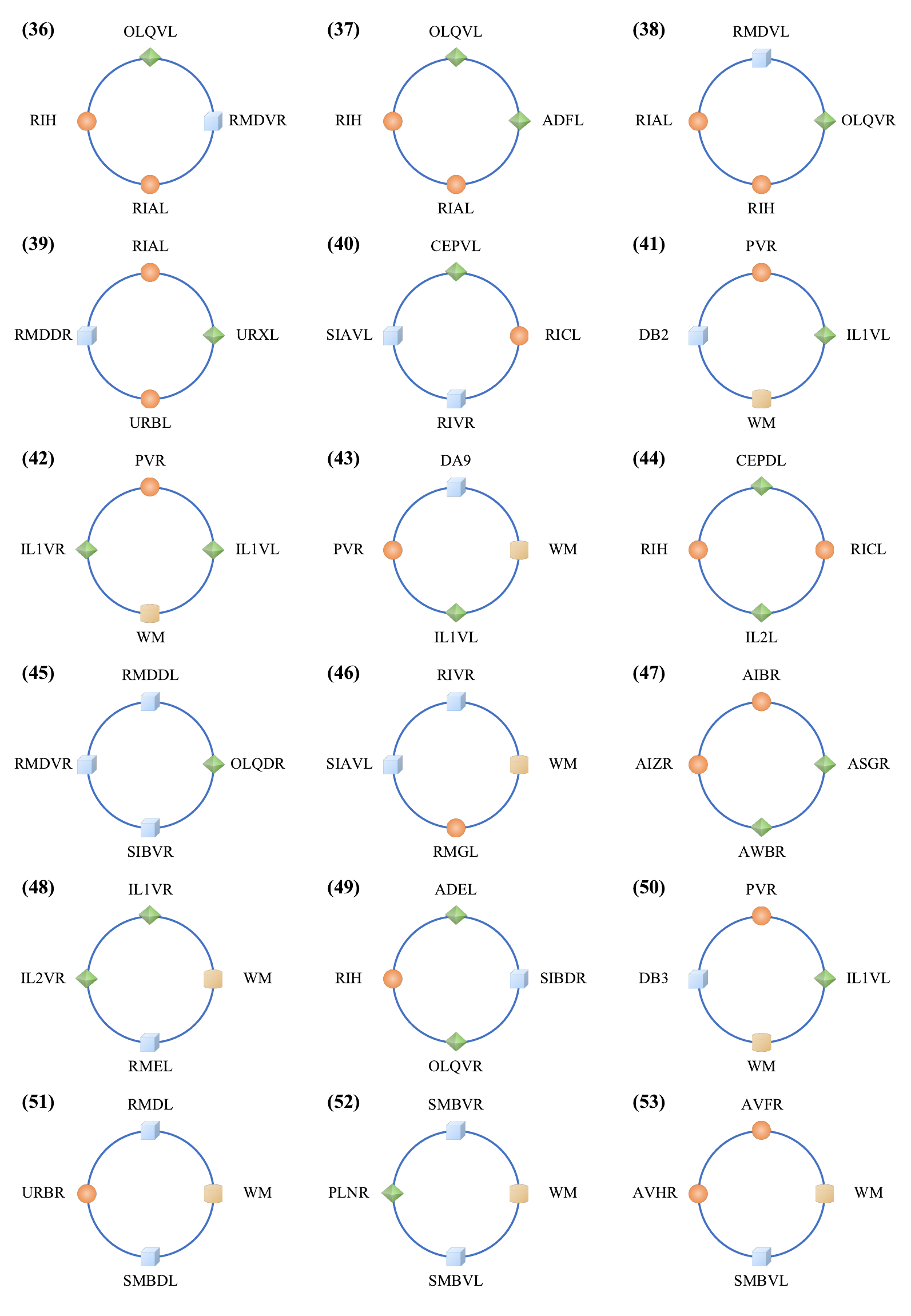}}
\end{figure}
\begin{figure}[H]
\centerline{\includegraphics[width=0.95\linewidth]{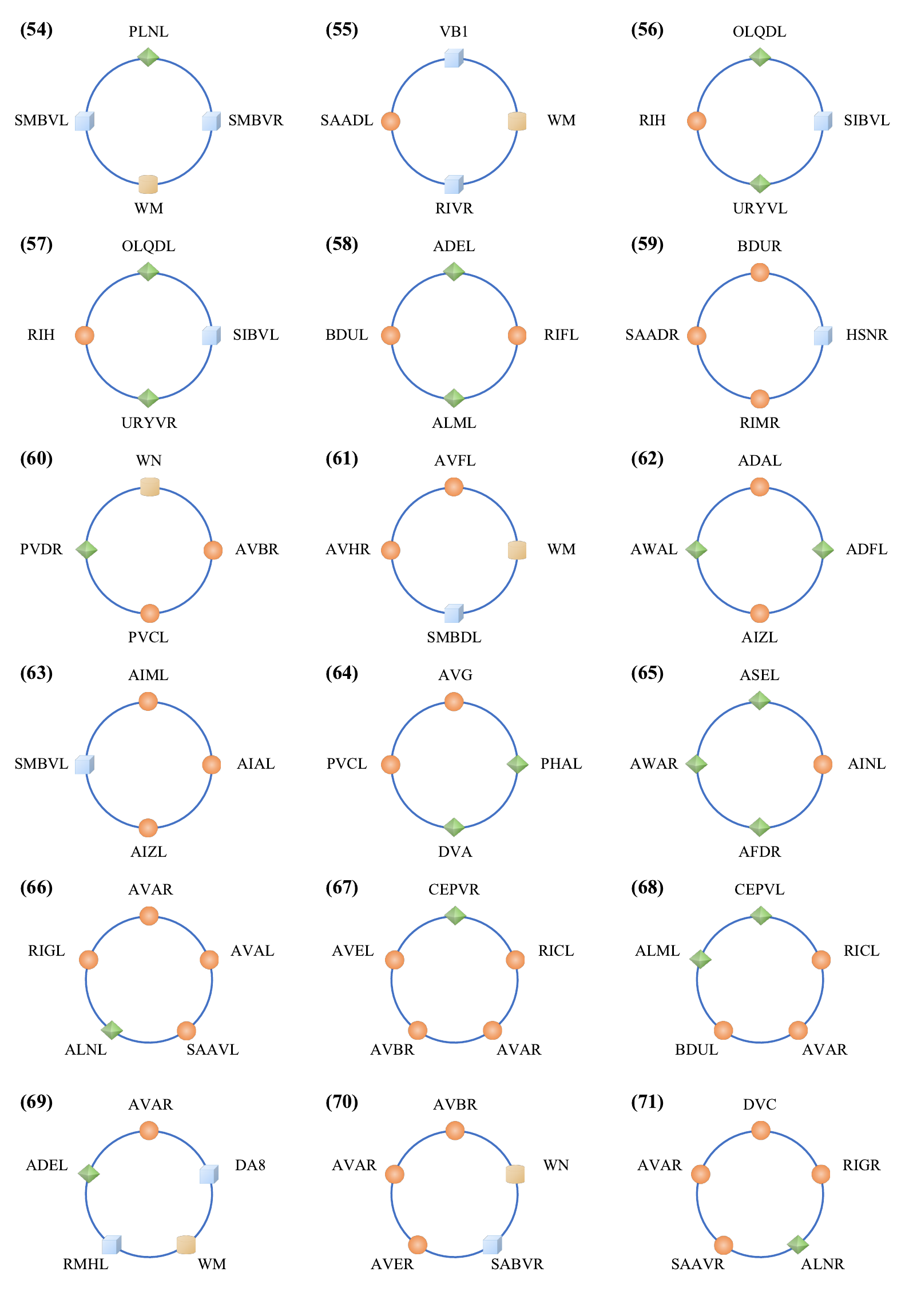}}
\end{figure}
\begin{figure}[H]
\centerline{\includegraphics[width=0.95\linewidth]{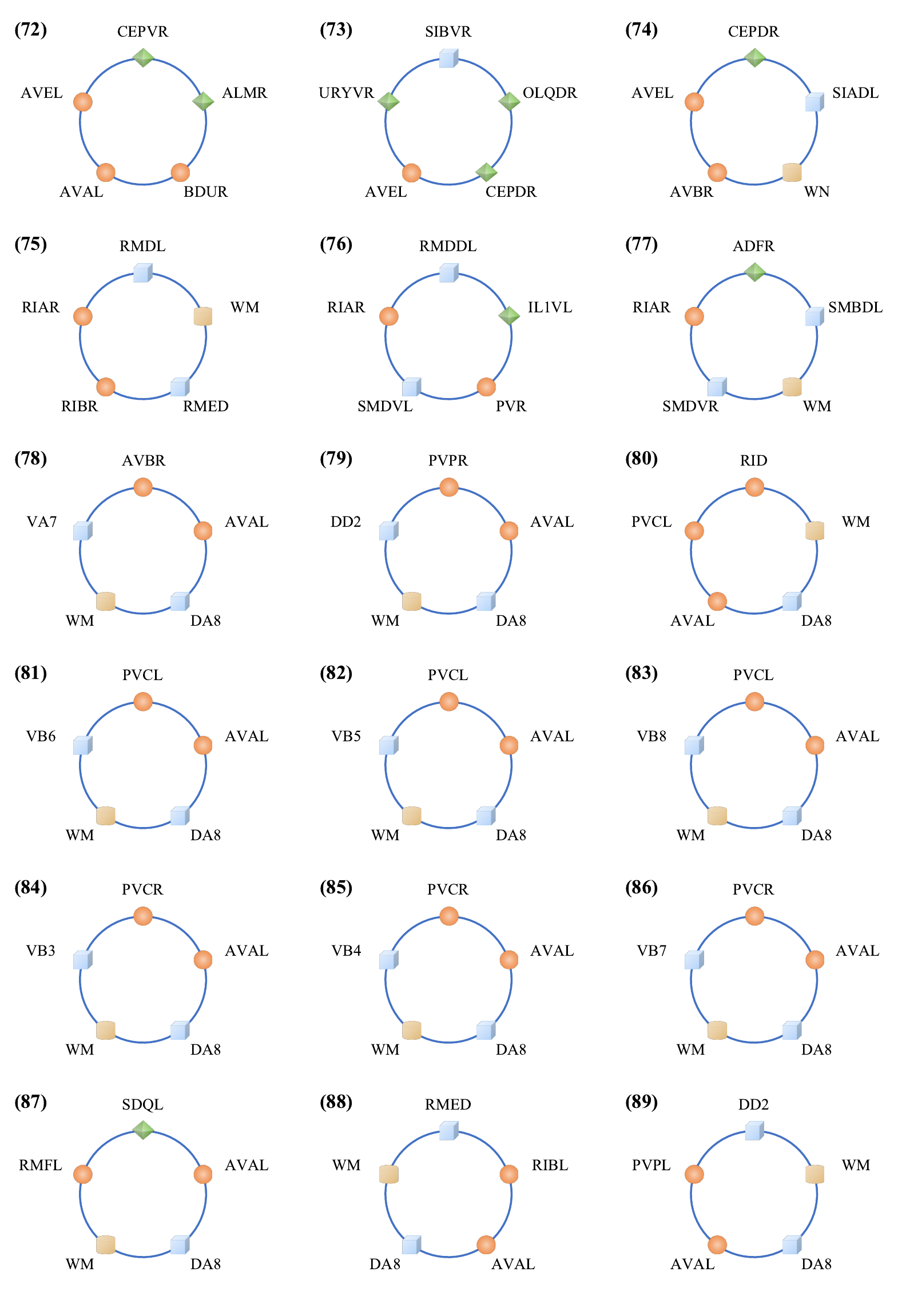}}
\end{figure}
\begin{figure}[H]
\centerline{\includegraphics[width=0.95\linewidth]{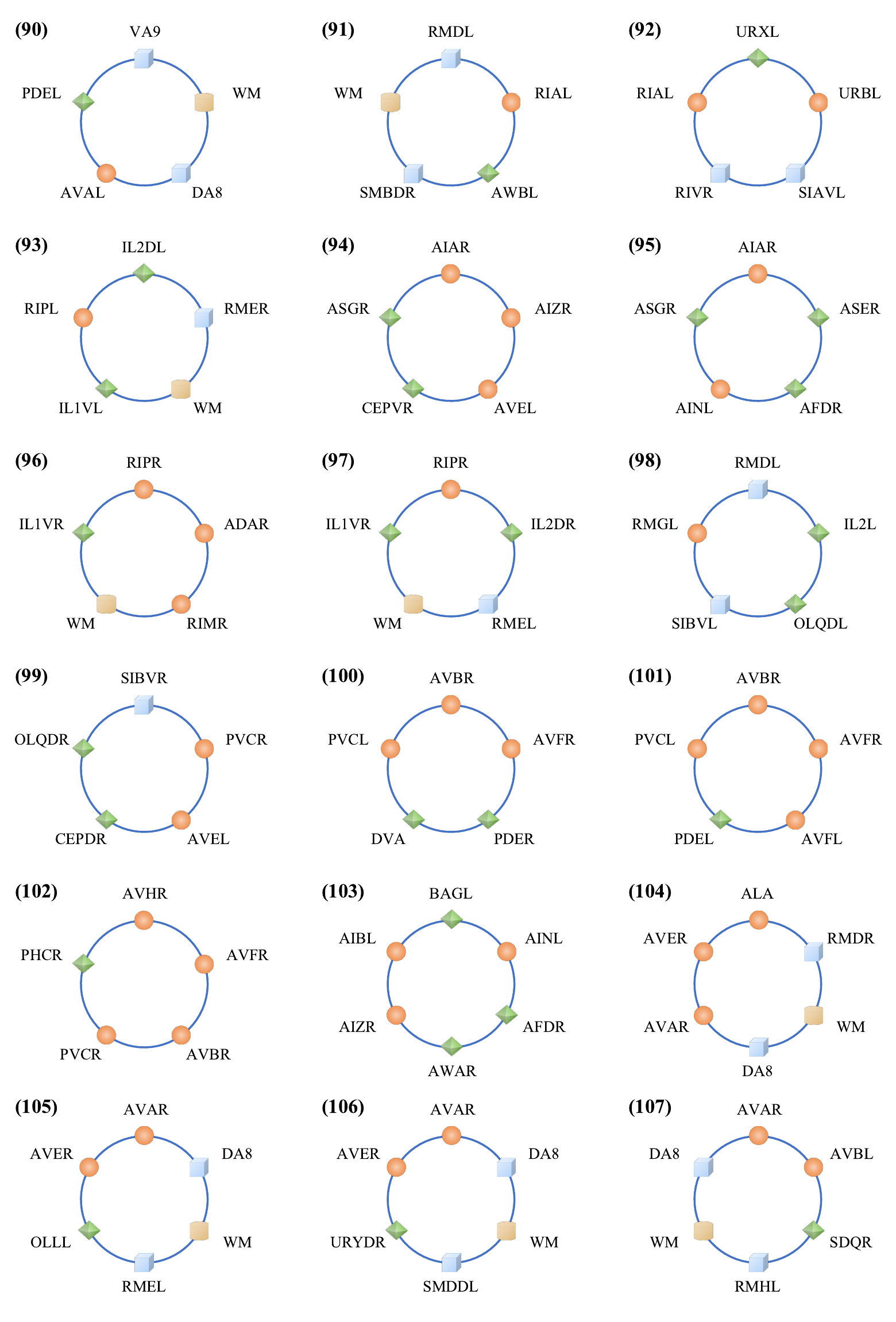}}
\end{figure}
\begin{figure}[H]
\centerline{\includegraphics[width=0.95\linewidth]{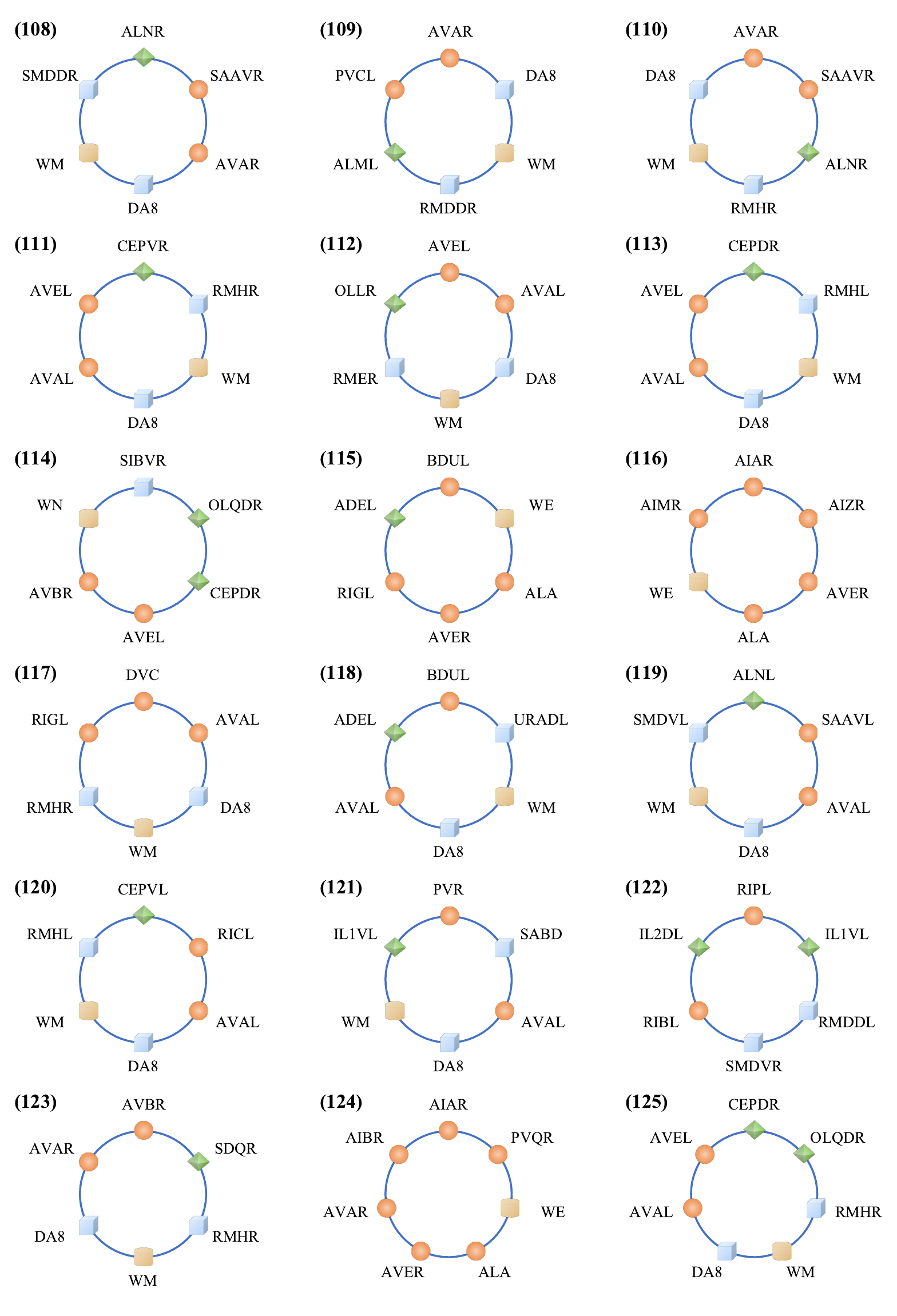}}
\end{figure}
\begin{figure}[H]
\centerline{\includegraphics[width=0.95\linewidth]{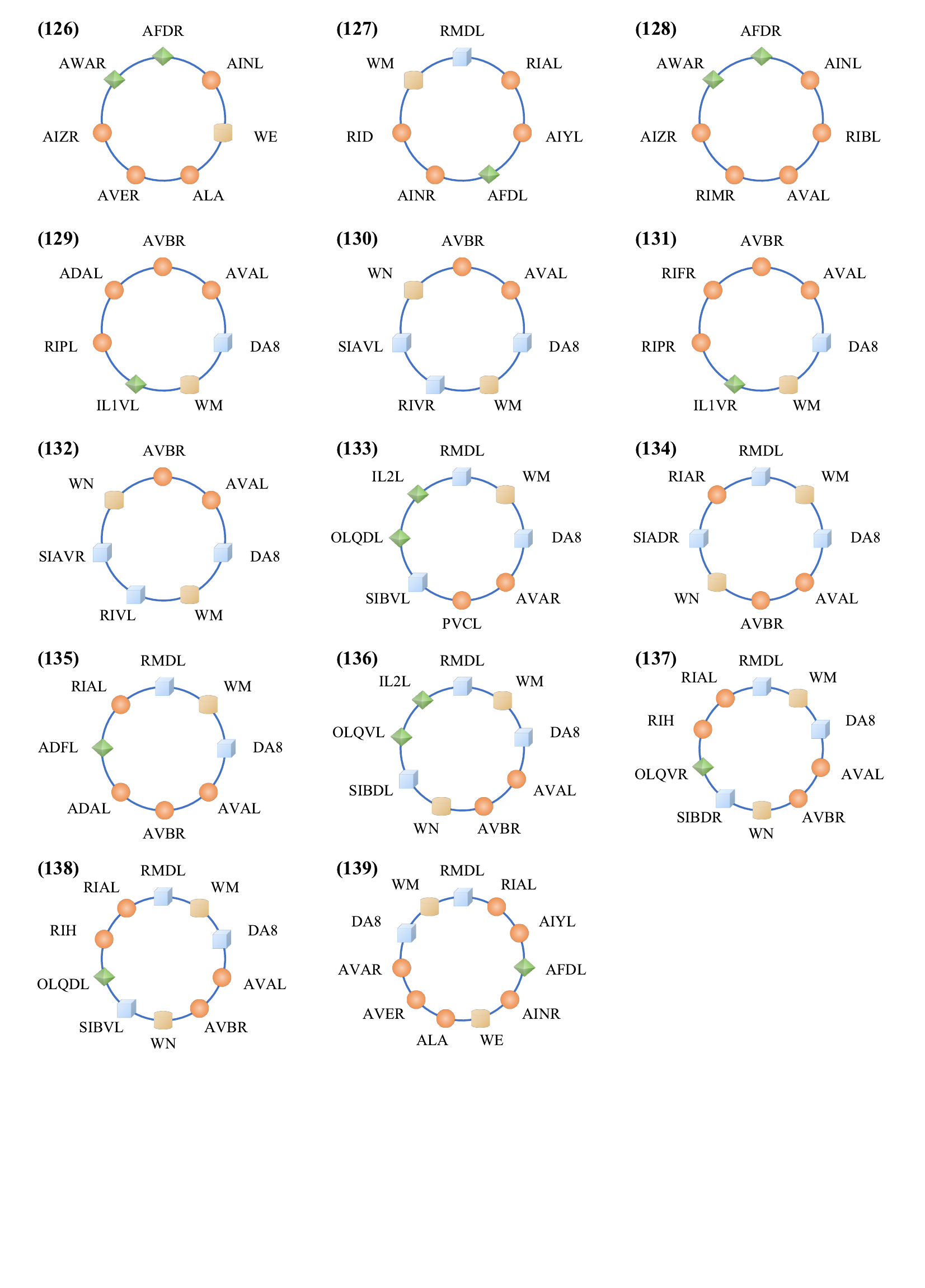}}
\end{figure}

\subsection*{2-cavity map}
\begin{figure}[H]
\centerline{\includegraphics[width=0.95\linewidth]{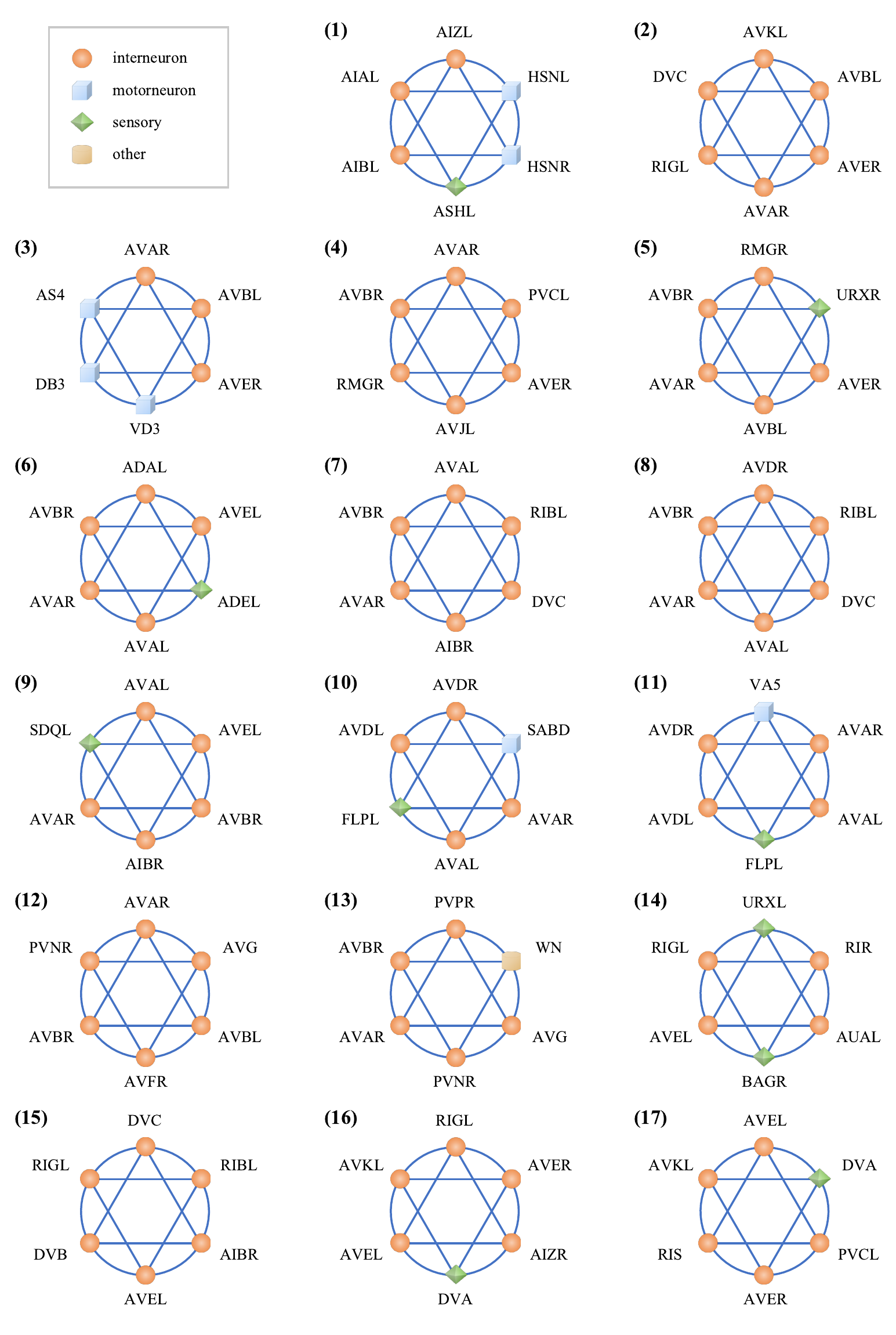}}
\end{figure}
\begin{figure}[H]
\centerline{\includegraphics[width=0.95\linewidth]{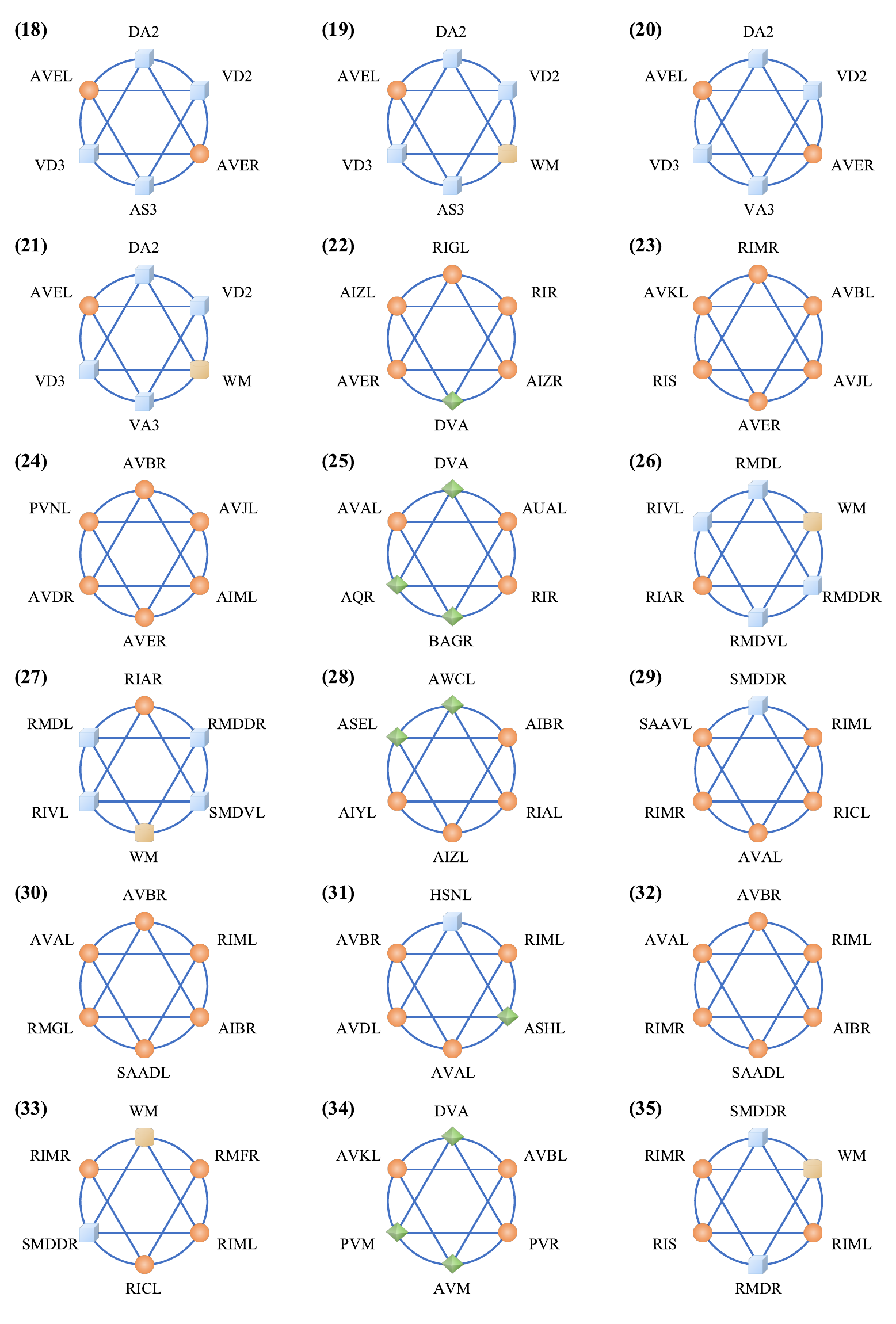}}
\end{figure}
\begin{figure}[H]
\centerline{\includegraphics[width=0.95\linewidth]{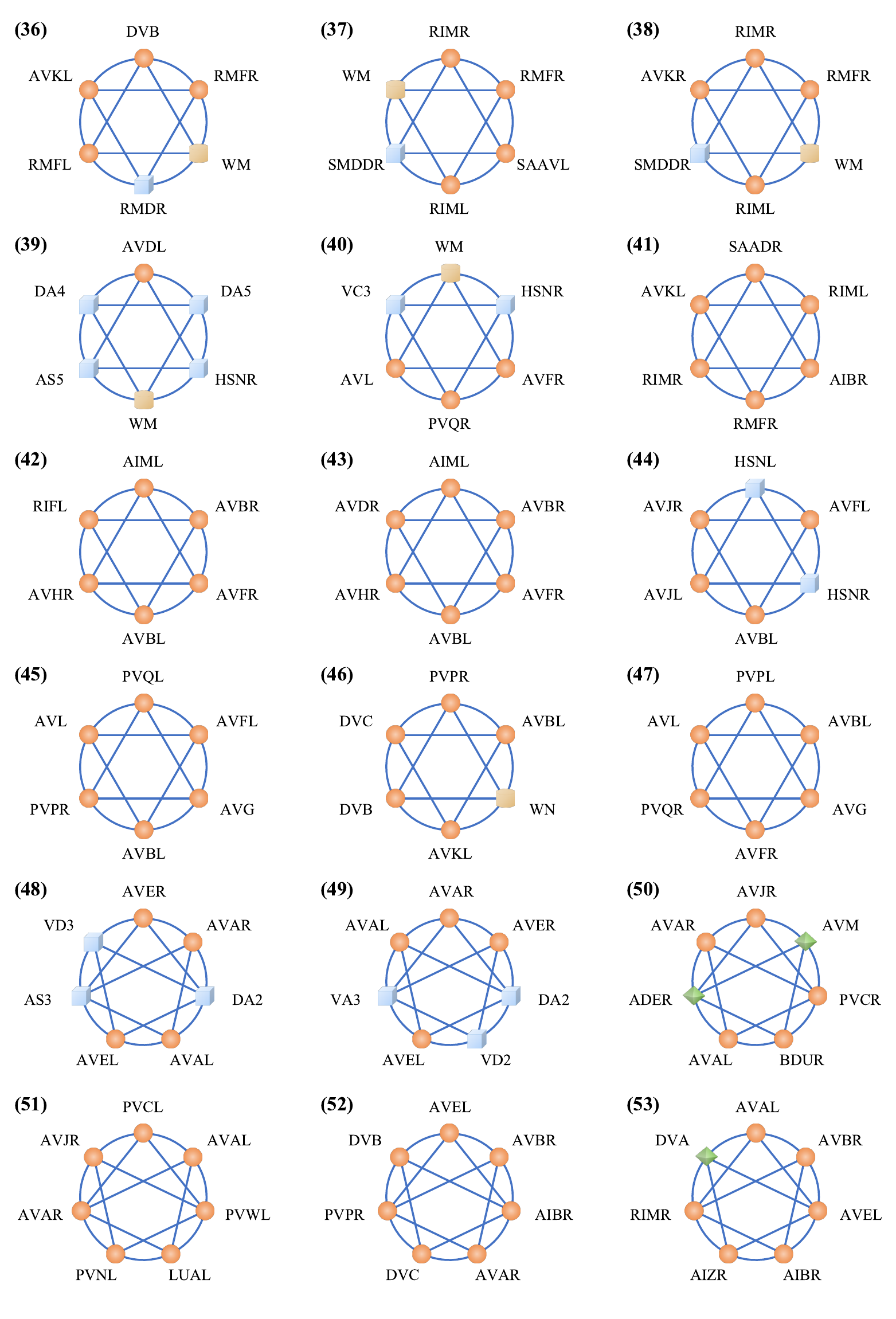}}
\end{figure}
\begin{figure}[H]
\centerline{\includegraphics[width=0.95\linewidth]{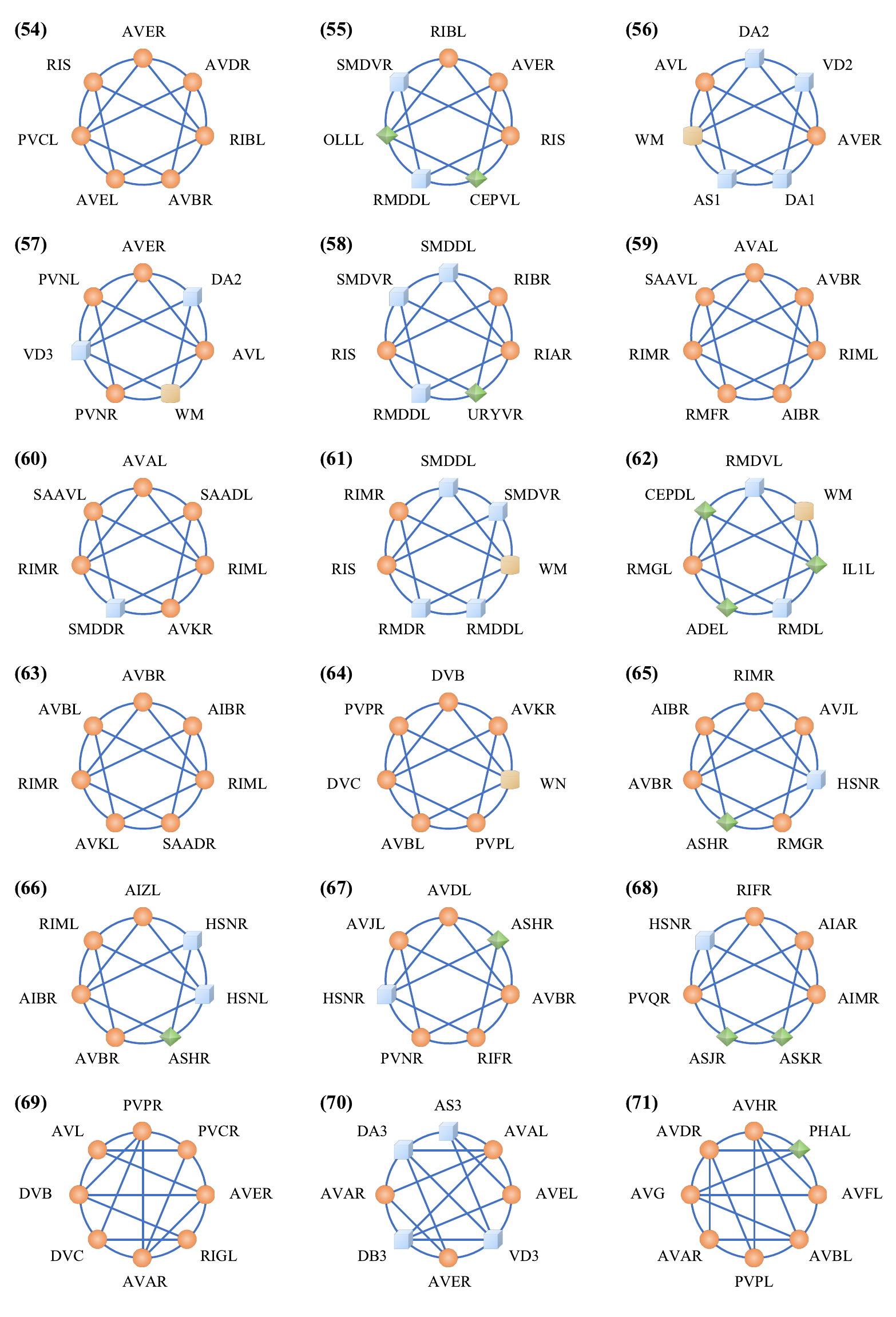}}
\end{figure}
\begin{figure}[H]
\centerline{\includegraphics[width=0.95\linewidth]{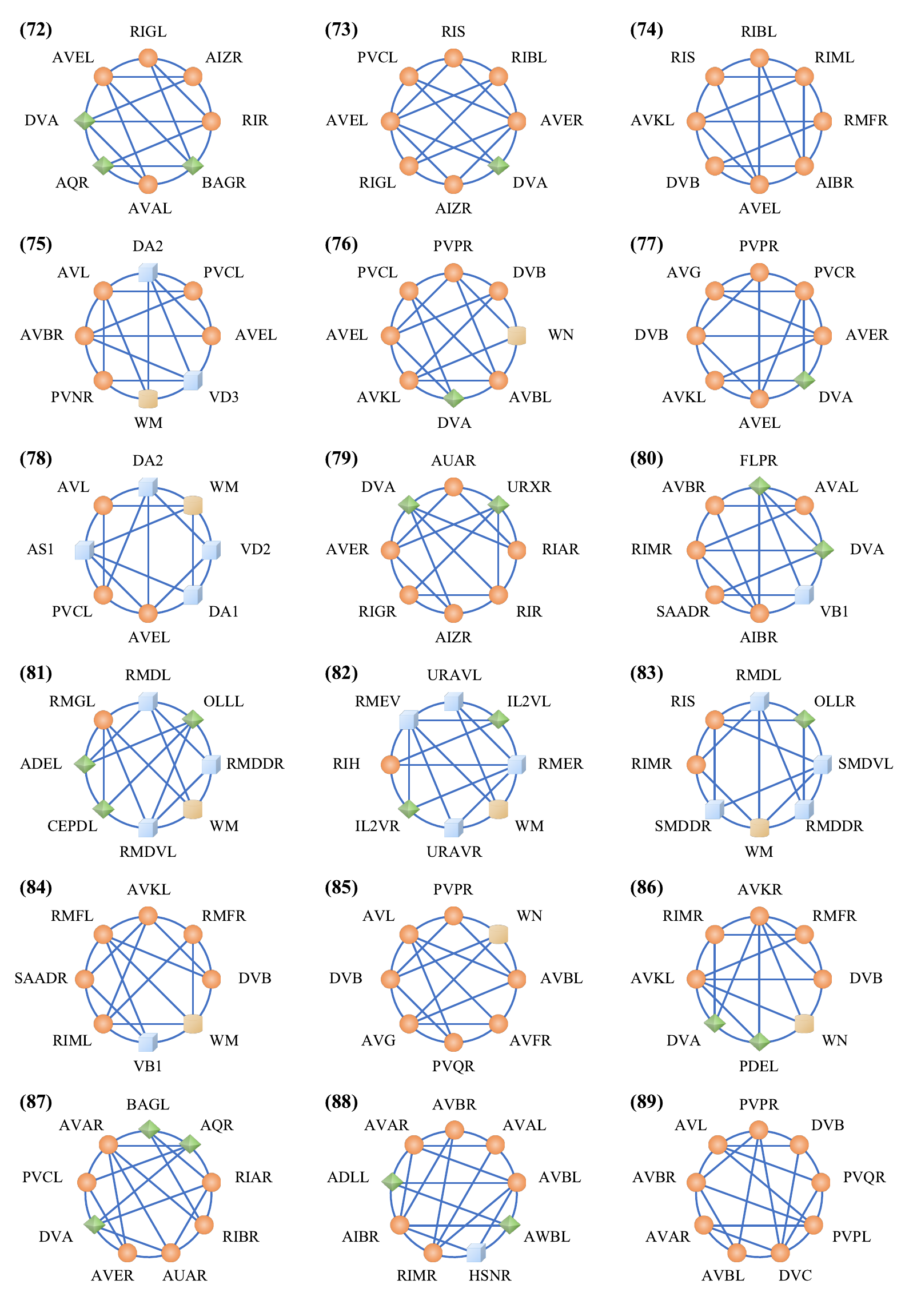}}
\end{figure}
\begin{figure}[H]
\centerline{\includegraphics[width=0.95\linewidth]{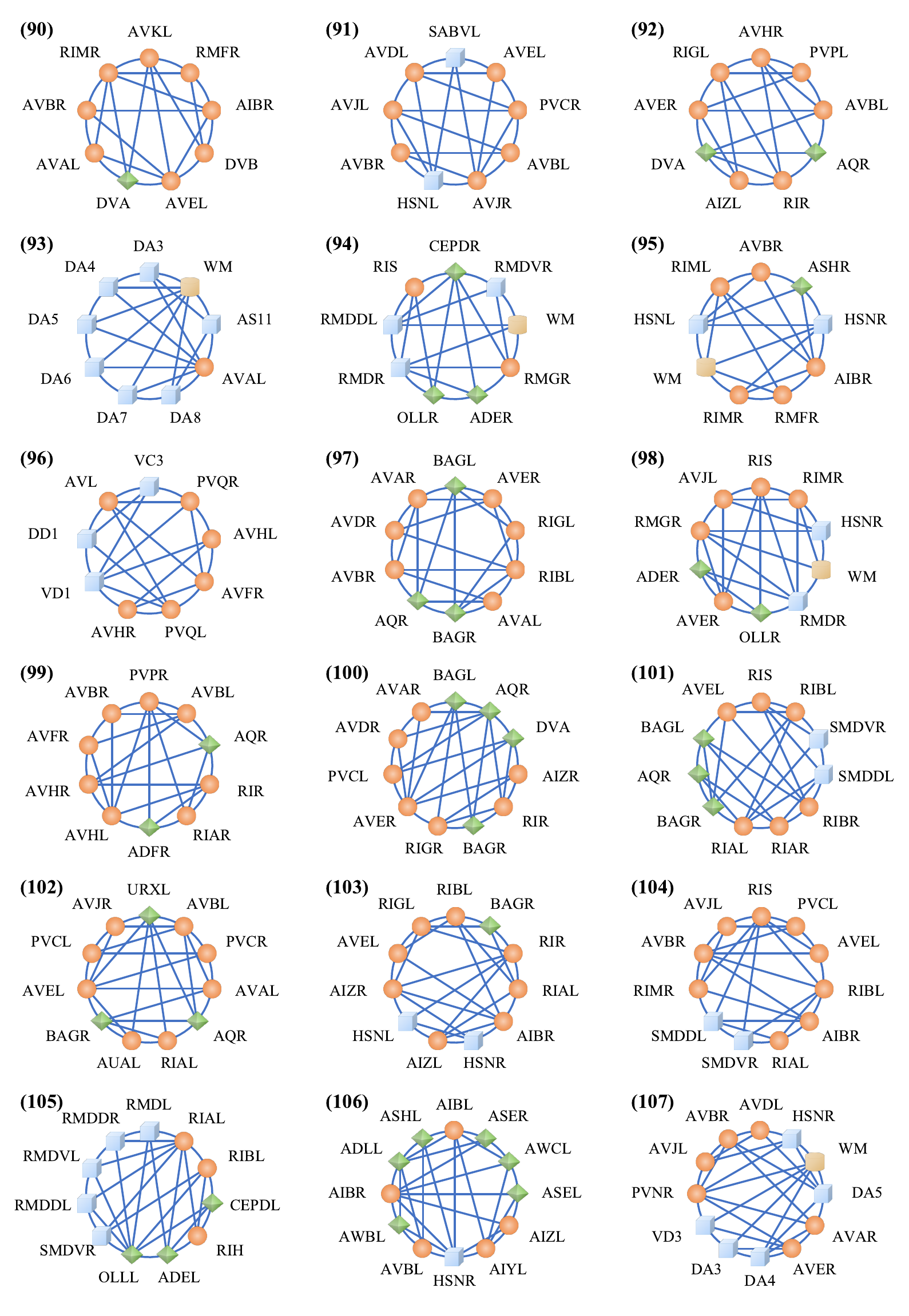}}
\end{figure}
\begin{figure}[H]
\centerline{\includegraphics[width=0.95\linewidth]{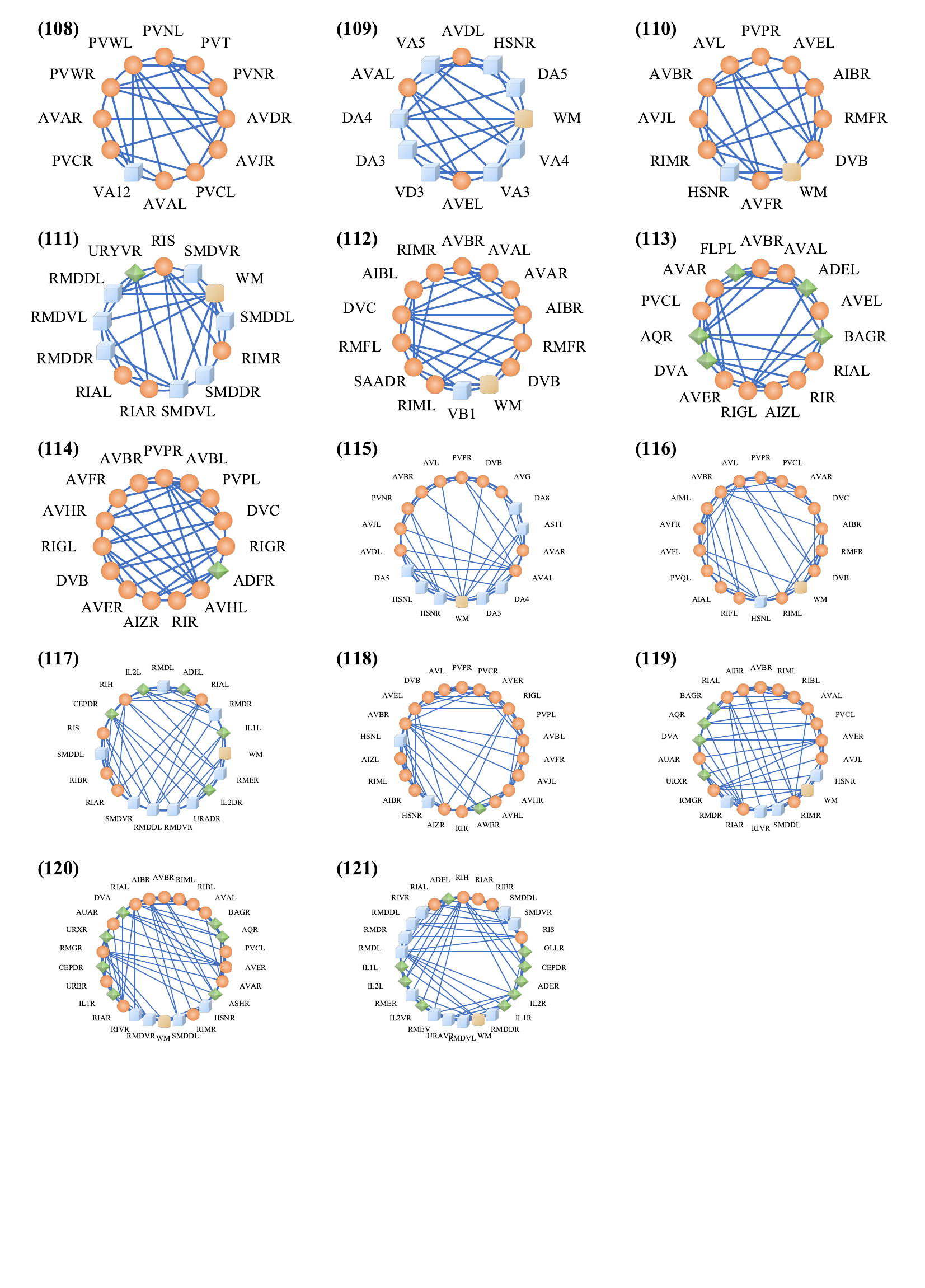}}
\end{figure}

\subsection*{3-cavity map}
\begin{figure}[H]
\centerline{\includegraphics[width=14cm]{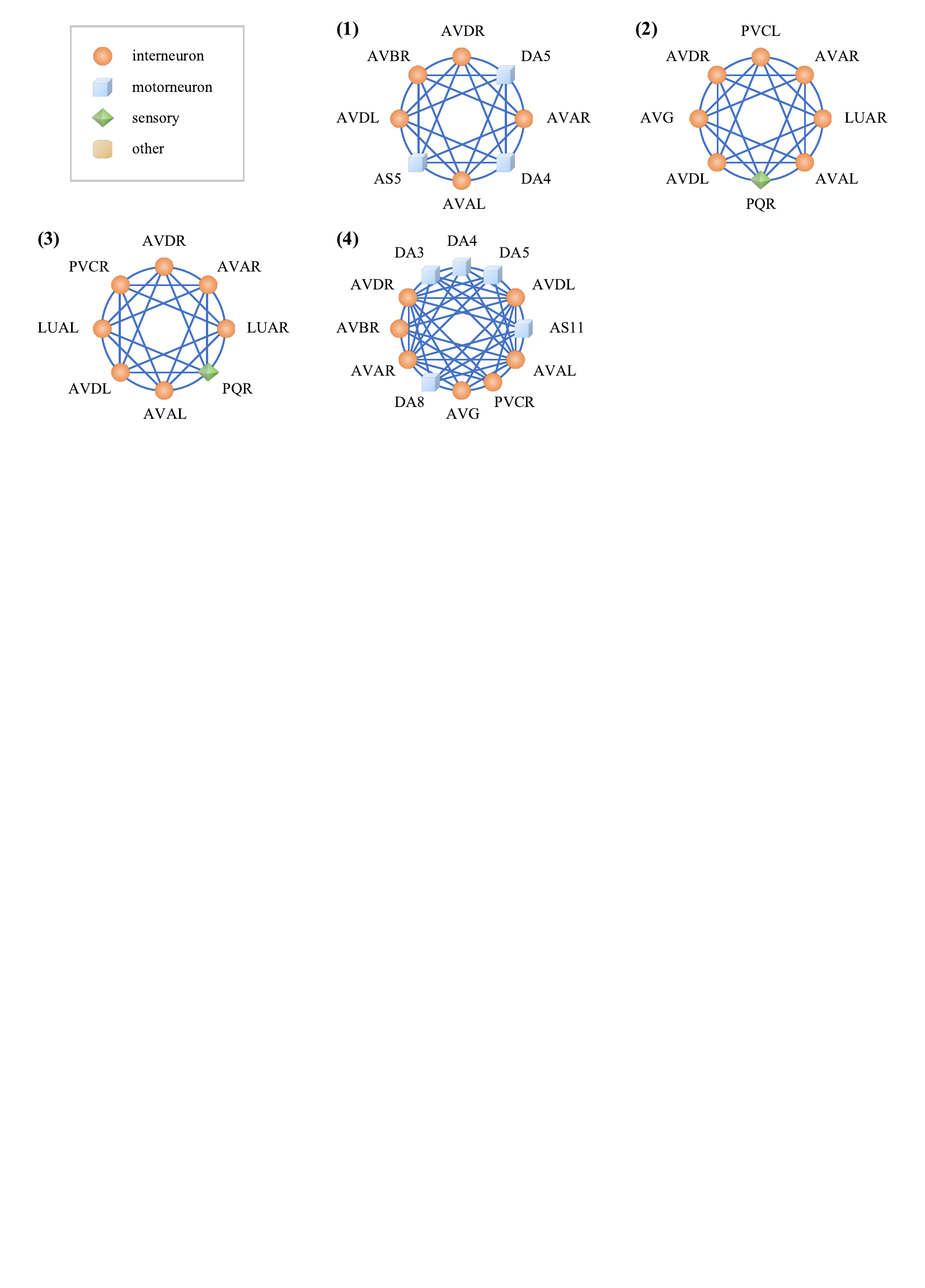}}
\end{figure}






\end{document}